\documentclass[lettersize,journal]{IEEEtran}

\usepackage{bbm}
\usepackage{url}
\usepackage{soul}
\usepackage{xspace}
\usepackage{amsthm}
\usepackage{booktabs}
\usepackage{hyperref}
\usepackage{makecell}
\usepackage{multirow}
\usepackage{graphicx}
\usepackage{ragged2e}
\usepackage{algorithm}
\usepackage{enumerate}
\usepackage{todonotes}
\usepackage{algpseudocode}
\usepackage{color, xcolor}
\usepackage{amsmath, amsfonts}

\newcommand{\ie}{\textit{i.e.,}\xspace}
\newcommand{\eg}{\textit{e.g.,}\xspace}
\newcommand{\etc}{\textit{etc.}\xspace}
\newcommand{\etal}{\textit{et al.}\xspace}

\newcommand{\figref}[1]{Fig.~\ref{#1}\xspace}
\newcommand{\tabref}[1]{TABLE~\ref{#1}\xspace}
\newcommand{\secref}[1]{Section~\ref{#1}\xspace}
\newcommand{\equref}[1]{Equation~\ref{#1}\xspace}
\newcommand{\agmref}[1]{Algorithm~\ref{#1}\xspace}

\newcommand{\toolname}{{\sc ScenTest}\xspace}

\begin{document}

\title{Practical, Automated Scenario-based \\ Mobile App Testing}

\author{Shengcheng~Yu,
        Chunrong~Fang,
        Mingzhe~Du,~
        Zimin~Ding,~
        Zhenyu~Chen,
        Zhendong~Su
\thanks{Shengcheng Yu, Chunrong Fang, Mingzhe Du, Zimin Ding, Zhenyu Chen are with the State Key Laboratory for Novel Software Technology, Nanjing University, China. \protect\\
E-mail: \{yusc, nandodu, dingzm\}@smail.nju.edu.cn, \{fangchunrong, zychen\}@nju.edu.cn}
\thanks{Zhendong Su is with the Department of Computer Science, ETH Zurich, Switzerland. \protect\\
E-mail: zhendong.su@inf.ethz.ch}
\thanks{Chunrong Fang is the corresponding author.}
}

\markboth{IEEE Transactions on Software Engineering}
{Shengcheng \MakeLowercase{\textit{et al.}}: Practical, Automated Scenario-based Mobile App Testing}

\maketitle

\begin{abstract}

The importance of mobile application (app) quality assurance is increasing with the rapid development of the mobile Internet. Automated test generation approaches, as a dominant direction of app quality assurance, follow specific models or strategies, targeting at optimizing the code coverage. Such approaches lead to a huge gap between testing execution and app business logic. Test scripts developed by human testers consider business logic by focusing on testing scenarios. Due to the GUI-intensive feature of mobile apps, human testers always understand app GUI to organize test scripts for scenarios. This inspires us to utilize domain knowledge from app GUI understanding for scenario-based test generation.

In this paper, we propose a novel approach, \toolname, for scenario-based mobile app testing with event knowledge graph (EKG) via GUI image understanding. \toolname tries to start automated testing by imitating human practices and integrating domain knowledge into scenario-based mobile app testing, realizing fully automated testing on target testing scenarios for the first time. \toolname extracts four kinds of entities and five kinds of corresponding relationships from crowdsourced test reports, where the test events and app GUI information are presented, and constructs the EKGs for specific scenarios. Then, \toolname conducts test generation for specific scenarios on different apps with the guidance of EKG with the combination consideration of app current state and testing context. We conduct an evaluation on \toolname on different aspects. The results show that the test generation of \toolname on the basis of EKG is effective, and \toolname reveals 150+ distinct real-world bugs in specific scenarios compared with representative baselines.

\end{abstract}


\section{Introduction}

\IEEEPARstart{M}{\textbf{obile}} 
\textbf{App Testing Dilemma.}
The number of mobile applications (app) has been increasing dramatically in recent years. The requirement for app quality assurance is getting demanding. It is challenging to ensure app quality when the apps are iterating rapidly. As a preliminary practice, app developers develop test scripts for apps based on specific testing frameworks. Such frameworks can execute the scripts \textit{exactly as recorded} \cite{behrang2019test, talebipour2021ui, xu2021guider} for mobile apps. However, such record and replay technologies rely heavily on the capabilities of app developers \cite{yu2021layout}, still costing a large amount of human labor. Automated testing \cite{baek2016automated} is a group of mainstream technologies for mobile app quality assurance. Automated app exploration technologies generate test cases automatically according to specific strategies, \eg random \cite{google2022monkey, machiry2013dynodroid}, model-based \cite{mao2016sapienz, su2017guided}, deep/reinforcement learning-based \cite{adamo2018reinforcement, pan2020reinforcement, romdhana2021deep}. However, whatever strategies the technologies take, they mostly take the code coverage as the optimization goal \cite{choudhary2015automated}. Therefore, such approaches ignore the business logic of the app under test during test generation, and they can hardly cover some significant but hard-to-reach testing scenarios.

\noindent \textbf{Perspective of Human Tester.}
Domain knowledge from human testers about the app testing scenarios is significant to make up for deficiencies of existing automated testing approaches. Experience can be learned from the testing cognition of human testers, who start app testing from the perspective of app GUI because mobile apps are GUI-intensive \cite{yu2021layout}. Human testers focus on the GUI widgets, the relationships among GUI widgets, and the operations applied to the widgets, which are important for automated approaches to understand. Also, test cases developed by human testers are organized according to testing scenarios and are closely related to app business logic, which inspires us to generate scenario-based test cases automatically. The prerequisite to better utilize the domain knowledge from human testers is to optimize the organization of such knowledge. Event knowledge graph (EKG) \cite{chen2020review} is an effective approach to organize information and obtain the links among different entities. Therefore, EKG can assist us in the scenario-based testing guidance.

\noindent \textbf{In this paper}, we propose a practical, automated \underline{\textbf{Scen}}ario-based mobile app \underline{\textbf{Test}}ing approach via image understanding and event knowledge graph, which is in short \toolname. \toolname has a comprehensive understanding of GUI image information with computer vision (CV) technologies, together with the textual information from crowdsourced test reports. Then it constructs the EKGs targeting at testing scenarios to guide the automated mobile app testing process with domain knowledge of app business logic from human testers.

\noindent \textbf{EKG Construction.}
Mobile app testing is a dynamic and procedural process. For the event-driven feature of mobile apps \cite{yu2021layout}, \toolname utilizes EKG instead of traditional KG to integrate the dynamic events and the static entities, together with the corresponding entity relationships. During the EKG construction, \toolname refers to crowdsourced test reports, which contain rich domain knowledge and testing procedures of human testers \cite{yu2021prioritize}. Crowdsourced test reports consist of app screenshots and textual descriptions \cite{tian2020furong}. Specifically, the textual descriptions include the reproduction steps, which describe the test events for specific testing scenarios For each test event in the reproduction steps, one app screenshot is assigned to intuitively show the app behaviors. To organize the information from crowdsourced test reports into EKG, \toolname decomposes the app screenshots and textual descriptions. For app screenshots, \toolname adopts CV technologies to extract GUI widgets, GUI structures, and existing texts; for reproduction steps in textual descriptions, \toolname extracts the operations and the corresponding objects of all events. Then, \toolname has a further understanding of the information from app screenshots. For example, the widget type is inferred with a convolutional neural network (CNN) model. In total, the EKG involves four entities (\ie Content, Widget, Operation, and Text) and five relationships (\ie TXT-TXT, CNT-OPT, CNT-TXT, CNT-WID, CNT-CNT), which are introduced in \secref{sec:relation} in detail. During the EKG construction, \toolname identifies entities from crowdsourced test reports, including operations, widgets, texts, \etc, and the relationships among different entities, like the extracted ``submit'' button from the app screenshot and the submit operation in the reproduction step. We use different test reports to complementarily construct the EKGs to avoid potential information lack in some test reports. Then, \toolname calculates the similarity among different entities and conducts the coreference resolution. Redundant information from different crowdsourced test reports is merged by linking different descriptions of the same entities, with the aim of supplementing inadequate information on the target scenarios.

\noindent \textbf{Scenario-based Mobile App Testing.}
At the beginning of the testing process, \toolname captures the app screenshot as the app current state. Then \toolname conducts semantic understanding of the app screenshot with the combination of traditional CV algorithms and deep learning models, including widget identification, GUI layout characterization, text extraction, \etc Widget identification refers to identifying all the existing widgets from the app screenshot, and the widget type, widget coordinates, and existing texts on the widgets are further extracted. Layout characterization refers to inferring the coordinate relationships among different widgets and analyzing the app activity structure. The aforementioned information helps form a nested app GUI structure file as a query to the EKG. EKG matches the app current state information with the entities and relationships in the EKG by similarity calculation and returns a widget list containing all the widgets in app current state and a probability for each widget. The probability is calculated based on similarities between each widget on current app GUI and the entities in the EKGs, and the probability indicates which widget can push forward the scenario-based testing if it is operated. After the widget operation, the new app current state will be analyzed and queried to the EKG. From the second step, existing operations will be queried to the EKG together with the app current state as a testing context sequence, to help better identify the next widget to be operated, until the end of the scenario-based testing. Considering the scenario diversity, we introduce the \textbf{\textit{Sub-Scenario}} concept, which refers to independent paths that can complete the target scenario. Sub-scenarios do not involve specific test inputs (\eg valid password or invalid password) but only indicate a path. \toolname applies such information into the tests during the concrete app exploration instead of integrating them into the testing scenario construction. During the EKG-guided test generation, \toolname adopts a depth-first strategy. For each test event that may trigger different sub-scenarios, \toolname can record such an event. When one sub-scenario is complete, the app state will be initialized to the state of the recorded event to start another sub-scenario, until all the sub-scenarios are complete.

\noindent \textbf{Empirical Evaluation}
An empirical evaluation is conducted to show the effectiveness of \toolname from five aspects: correctness, reliability, accuracy, adequacy, and usefulness. The evaluation is conducted on the eight most widely used testing scenarios and based on a dataset of 124 mobile apps. The results show that \toolname can effectively construct the event knowledge graphs for the testing scenarios, and the automated scenario-based testing is accurate and adequate. Moreover, compared with the representative baselines, \toolname can effectively find 150+ distinct real-world bugs in specific scenarios on different mobile apps.

We declare the following noteworthy contributions:

\begin{itemize}
	\item We propose a novel framework that utilizes image understanding and event knowledge graphs for scenario-based mobile app testing.
	\item We propose a novel approach to automatically construct event knowledge graphs for mobile app testing scenarios from crowdsourced test reports.
	\item We propose a novel approach to guide the automated mobile app testing with EKG targeting at scenarios from the perspective of human testers.
	\item We design and implement a tool, \toolname, and conduct a large-scale experiment to evaluate its effectiveness from different aspects.
\end{itemize}

More information and the reproduction package are available at: \underline{\url{https://zenodo.org/records/11118420}}. \vspace{0.1cm}

The rest of the paper is organized as follows. In \secref{sec:example}, we provide an illustrative example to show the limitations of current approaches and the workflow of \toolname. In \secref{sec:ekg}, we elaborate on the detailed approach regarding the EKG design. In \secref{sec:test}, we illustrate the detailed approach with regard to the scenario-based test generation process. In \secref{sec:exp}, the empirical evaluation is presented. In \secref{sec:dis}, a discussion on the features and application scope of \toolname is proposed. \secref{sec:rw} presents the related work from two aspects, the KG application in software engineering and the image-aided software testing. Finally, in \secref{sec:con}, the conclusion is made to the whole paper.

\section{Illustrative Example}
\label{sec:example}

\begin{figure*}[!htbp]
\centering
\includegraphics[width=\linewidth]{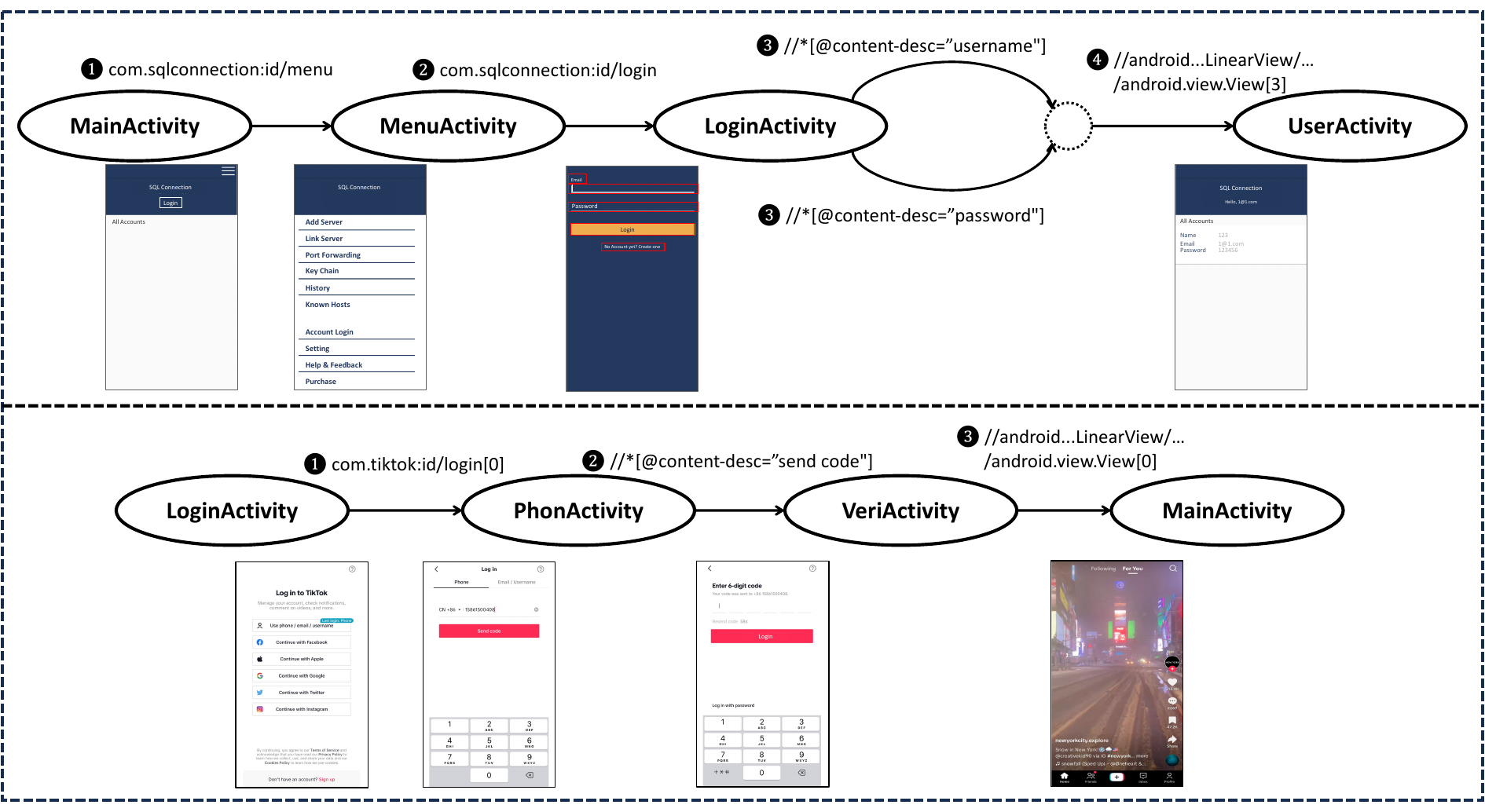}
\caption{Illustrative Example: Login Scenario Process in Two Different Crowdsourced Test Reports}
\label{fig:exreport}
\end{figure*}

To illustrate the motivation of this paper, we provide an example to depict the challenges. \figref{fig:exreport} depicts the process of testing the login scenario on a database management app and a short video app, respectively. The screenshots are from crowdsourced test reports and correspond to the textual descriptions below. There are several steps in different crowdsourced test reports.

\textbf{Limitations of existing automated approaches.}
This login process is an easy business logic for human testers to test, while it is hard to automatically generate such test events even for the state-of-the-art automated testing approaches. We run three approaches for two hours each to conduct an investigation: Monkey \cite{google2022monkey}, a widely used random approach provided by Google, Stoat \cite{su2017guided}, a state-of-the-art model-guided approach, and UniRLTest \cite{zhang2022unirltest}, the most advanced approach that adopts reinforcement learning technology to explore app states. Then, we review all the generated test events, and all three approaches fail to completely cover the login scenario in the test events\footnote{the correct username and password strings are provided as samples}. The possible reason may be that the app supports some other business logic without being logged in, which may trap the testing execution, or the login process can be interrupted before the whole scenario is fully completed.

Existing automated testing approaches construct models of the AUTs following specific strategies. Such models include the app activity information and the transition information, which only indicate the transition relationships among different apps, while the business logic is not obtained from the exploration process, thus making it hard to understand the scenario to be tested. Consequently, it is hardly possible for current automated testing approaches to generate test events considering business logic, and it is challenging to generate scenario-based test events. Without the guidance of human knowledge, some critical parts of the app involving complex business logic will be missed by existing testing approaches. Therefore, we believe it is necessary to introduce the domain knowledge of human testers into the testing process as effective guidance.

The test migration approaches, as another group of approaches, may seem to be similar to \toolname from the execution perspective. However, the design purpose of \toolname and test migration approaches are actually quite different. Test migration tools take source test event sequences developed by developers or extracted from the source apps as a reference, while \toolname adopts an exploration strategy, using the EKGs to guide the exploration of the target scenarios on different apps. Test migrations tools are designed to strictly migrate the test event sequences to target apps to accomplish pre-defined test event sequences, while \toolname is designed to explore the specific target scenarios of apps with the EKG guidance and to be more flexible when trying to complete the testing on specific scenarios. Different test event sequences (\ie sub-scenarios) are likely to be covered during the exploration. Since it is indeed true that our approach has a completely different purpose from test migration tools, we do not conduct corresponding experiments with test migration tools.

\textbf{EKG Construction of the Illustrative Example.}
\toolname constructs an event knowledge graph for the Login scenario (not a complete one in our evaluation) based on crowdsourced test reports, which integrate domain knowledge from human testers into the automated exploration. The app screenshots provided in the crowdsourced test report are shown in \figref{fig:exreport}, and the corresponding textual descriptions are as follows:

\begin{center}
\setlength{\fboxrule}{1pt}
	\fbox{
	\parbox{0.9\linewidth}{
	\textbf{\underline{Report 1}: When the app launches, click on the menu button, choose the ``Account Login'' option, then click on the Login button, then input the username ``admin'' and password ``123456'', and then click on the Login button. The page is turned to the user info page, while it shows another account ``123'' with email ``1@1.com''.} \\
	\textbf{\underline{Report 2}: Click on the ``Use phone/email/username'' button to choose a specific login method, and then input the phone number in the textfield and click on the ``Send code'' button. Then input the code to login, and after clicking on the Login button, the app automatically turns to the main activity.}
	}
}\end{center}

\begin{figure*}[!htbp]
\centering
\includegraphics[width=\linewidth]{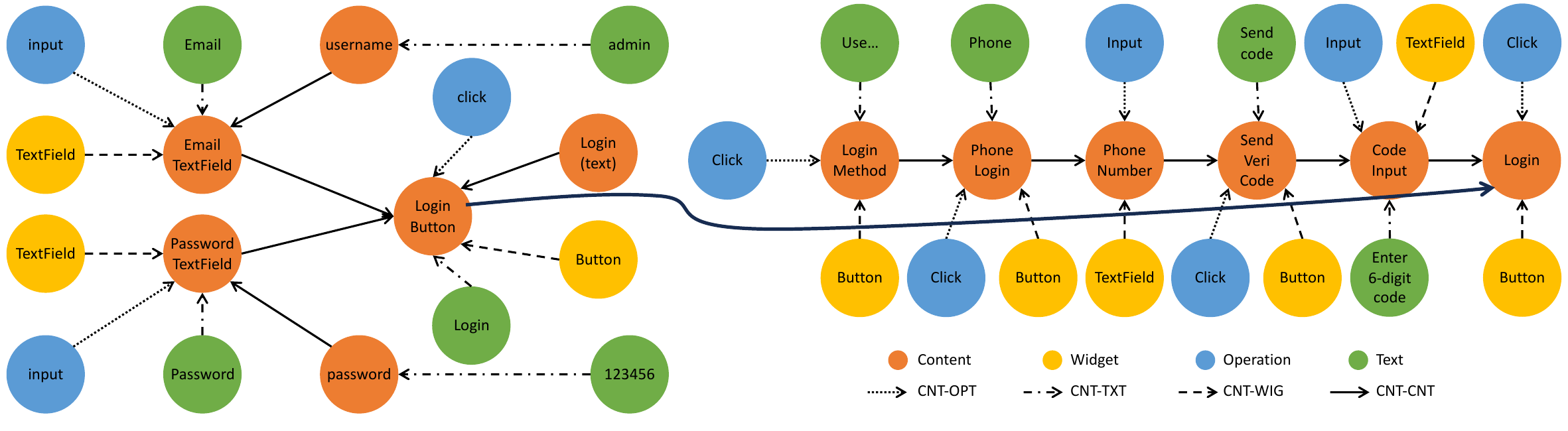}
\caption{Illustrative Example: Event Knowledge Graph}
\label{fig:exkg}
\end{figure*}

The first step is to decompose the app screenshots and textual descriptions in each test report. Take the first report as an example, for the sentence ``\textit{then input the username `admin' and password `123456', and then click on the Login button}'', \toolname can extract entities as \textit{[input, username, ``admin'', password, ``123456'', click, ``Login'', button]}, and for the app screenshot, \toolname can extract the existing widgets attached with texts and identify the widget type with a DL model. Therefore, for the screenshot of the third step (as the red rectangles shown in \figref{fig:exreport}), \toolname can extract texts including ``email'', ``password'', ``login'', ``No Account yet? Create one'' and widgets including two \texttt{TextField} and a \texttt{Button}. Then, the entities are extracted: \textbf{\underline{CNT}}: ``Email TextField'', ``Password TextField'' (from image), ``Login Button'' (from image), ``admin'', ``username'', ``123456'', ``password'' (from text), ``Login'' (from text); \textbf{\underline{WID}}: ``TextField'', ``Button''; \textbf{\underline{OPT}}: ``click'', ``input''; \textbf{\underline{TXT}}: ``Email'', ``Password'', ``Login''. The processing is similar to the second crowdsourced test report in the illustrative example.

\toolname then obtains the relationships for different entities. The relationships in the illustrative example are shown in \figref{fig:exkg}. In this example, five kinds of relationships are involved. Specifically, one thing to notice is that the \textbf{\underline{TXT-TXT}} relationship is designed for the processing of different crowdsourced test reports, which is labeled with a bolder line in \figref{fig:exkg} between the two crowdsourced test reports. In order to make the EKG more complete, \toolname absorbs different descriptions on the same objects or synonyms used in different reports, \eg ``login'' and ``sign in''. Such texts are processed in the coreference resolution part during the EKG construction. Coreference resolution also processes similar entities among different crowdsourced test reports (like the entities linked with the bold line in \figref{fig:exkg}). ``Login Button'' refers to the concrete button that appears on the app GUI, ``Login (text)'' refers to the texts extracted from the textual descriptions from crowdsourced test reports, and ``Login'' refers to the texts extracted from the app screenshots from crowdsourced test reports. This can help match the app screenshots and textual descriptions within the same crowdsourced test report. Regarding the relationship links between ``Login Button'', ``Login (text)'' and ``Login'', we do not link the ``Login (text)'' and ``Login''. This is due to a sequence issue. When the relationship between ``Login (text)'' and ``Login Button'' and the relationship between ``Login'' and ``Login Button'' are constructed, the relationship between ``Login'' and ``Login (text)'' will not be constructed to avoid the loop. The loop may bring circular queries, which may bring extra time overhead \cite{zhao2021brain} to confirm the coreference to the entity. Therefore, during the EKG construction, we try to avoid forming loops in the EKGs. For the same scenario in different crowdsourced test reports, we extract different sub-scenarios. For the first report, the sub-scenario for the login scenario is through the combination of username and password. For the second report, the sub-scenario for the login scenario is through the combination of phone number and verification code. However, both of the two reports end the scenario-based testing by clicking on the ``Login'' button. The commonality between these two reports, which is to click on the ``Log in'' button on different apps, is identified and linked in the coreference resolution.

\begin{figure}[!htbp]
\centering
\includegraphics[width=\linewidth]{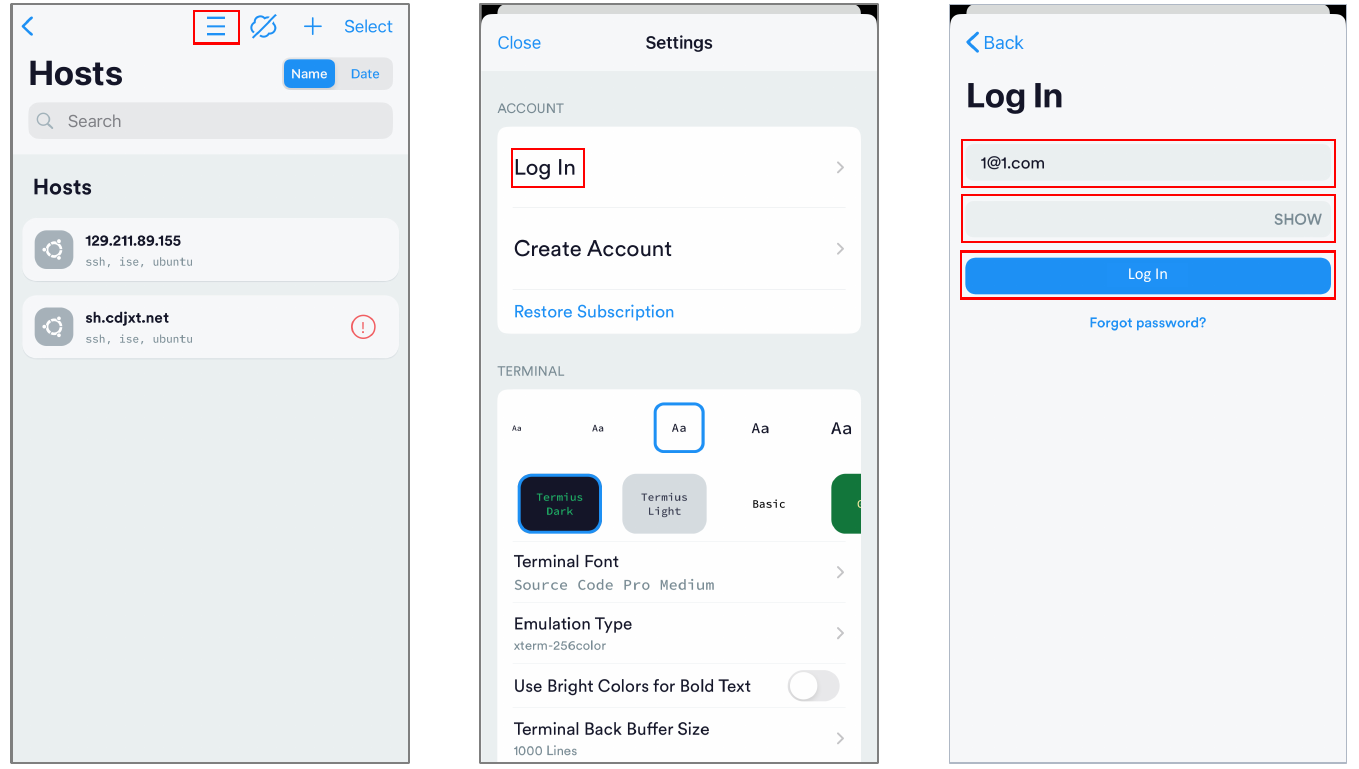}
\caption{Illustrative Example: Test Generation}
\label{fig:exgen}
\end{figure}

\textbf{EKG-Guided Test Generation of the Illustrative Example.}
Then, the EKG can be used to guide the test generation for a different app. Generally, \toolname will maintain a memory that records the context of the testing events. First, \toolname will query the EKG with a null memory and the current app screenshot to indicate the start. The current app screenshot is analyzed to extract widgets and existing texts, like the widget indicating ``menu'' in the first image in \figref{fig:exgen}. Then, the EKG returns a widget list, attached with a probability for each widget that indicates whether the widget should be operated. The probabilities are calculated as the similarity between widgets extracted from current app activity and the \textbf{\underline{CNT}} and \textbf{\underline{TXT}} entities in the EKG. The calculation process of the similarity that determines the widget operation is presented in \secref{sec:coreference}. A more detailed process of how to determine the probability is presented in \secref{sec:case}. In this example, the menu icon is assigned with the highest probability. This is due to the presence of ``click on the menu button'' presence in Report 1, and it is a pre-step of login action. Therefore, when the widget indicating ``menu'' shows in the new app, the corresponding widget will be matched with the entities extracted from Report 1, which is omitted in \figref{fig:exkg} in order to highlight the key steps in a clearer way so we only present part of the EKG from Report 1 and 2. Actually, the menu widget is also present in many other reports. Therefore, as a typical pre-step of login operations, it will be assigned with a high probability. When the EKG returns the results, the \textbf{\underline{OPT}} is also returned, which is \textit{click} in this step. Then, such a test event is recorded in the memory. The new memory and the new app activity screenshot are formed as a new query to the EKG. At the last step, which is clicking on the ``Log in'' button, an \textit{end} signal will be accompanied to indicate the end of the test generation.

In this section, we hope to show the limitations of existing approaches by a pilot comparison study. We run the existing automation tools, and the results show that they may reach relatively good code coverage, but they fail to complete the Login scenario even if they are provided with valid username and password. We also provide the analysis on possible reasons. We believe that these automated tools cannot complete other scenarios, which are even more complex, let alone traverse all possible sub-scenarios. Our approach is therefore proposed to deal with such a situation. \toolname is designed to explore the specific scenarios of AUTs under the guidance of EKGs, instead of reaching high coverage but ignoring the business logic. \figref{fig:exreport}, \figref{fig:exkg}, and \figref{fig:exgen}, as a whole, are presented to elaborate how \toolname completes the specific testing scenario with the guidance of the EKG. \figref{fig:exreport} shows the crowdsourced test reports as the EKG construction data source, \figref{fig:exkg} shows the constructed EKG with these two reports (cot the complete one in our experiment for Login scenario), and \figref{fig:exgen} shows how \toolname pushes forward the test faced with a new app with the Login scenario. We hope to illustrate the advantage of \toolname in terms of focusing on one specific testing scenario according to the business logic whereas existing automated tools cannot accomplish. Also, through the illustrative example, we hope to elaborate how different crowdsourced test reports are separately analyzed and then merged by the coreference resolution, how the information extracted from crowdsourced test reports is used to construct the EKGs, and how the EKGs are used to guide the scenario-based testing on a completely different new app. This helps claim the generalizability of the EKG. This example, together with the pilot comparison study, shows our motivation to propose \toolname to cover the testing scenarios to generate understandable tests.

\begin{figure*}[!htbp]
\centering
\includegraphics[width=\linewidth]{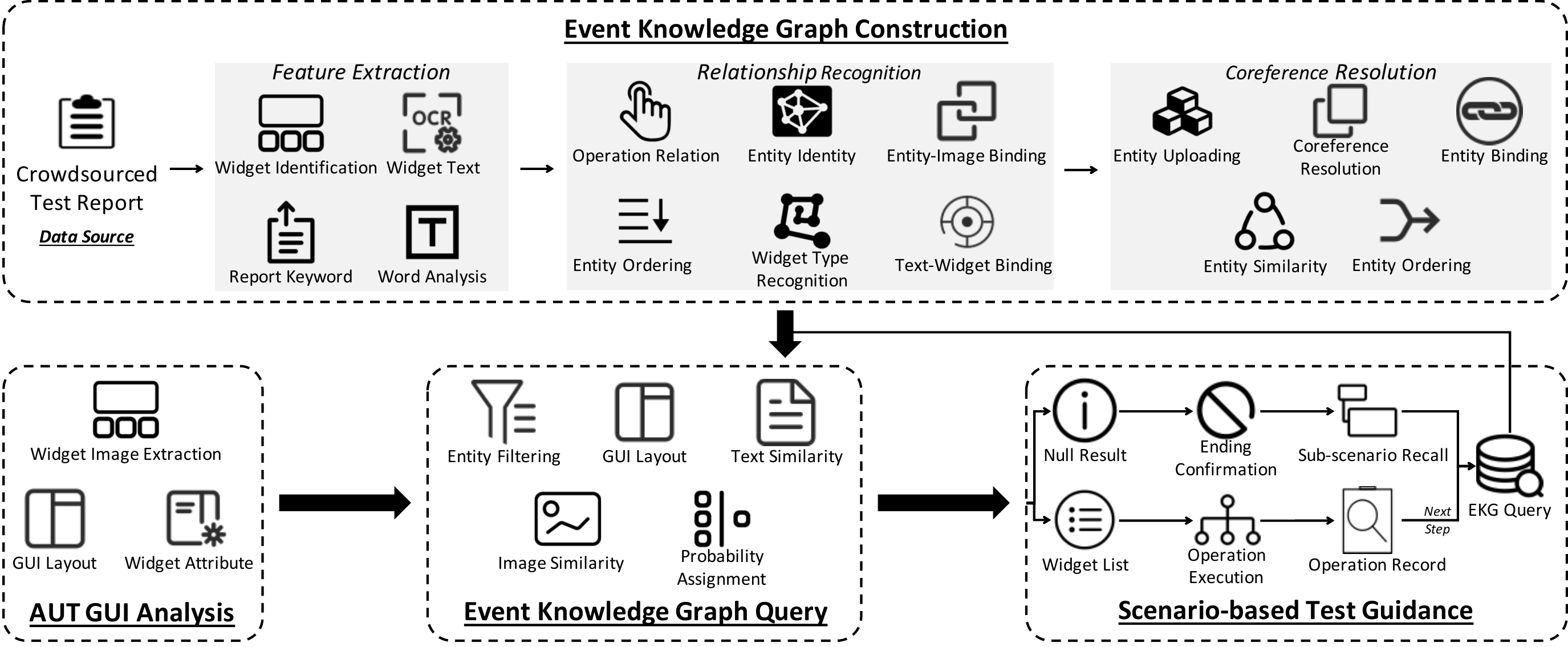}
\caption{\toolname Framework}
\label{fig:framework}
\end{figure*}

\section{EKG Construction}
\label{sec:ekg}

The event knowledge graph (EKG) is a kind of effective way to store the domain knowledge of mobile app testing \cite{guo2020crowdsourced}. The data source to construct EKG for testing scenarios is the crowdsourced test reports \cite{gao2019successes}. Crowdsourced test reports contain manual test events that present the detailed steps of operations and the target widgets, and app screenshots of such steps are attached. The EKG construction is shown in \figref{fig:framework} (the top half) and \agmref{alg:construction}. First, the textual descriptions and app screenshots are automatically analyzed, and \toolname extracts corresponding entities for testing scenarios (\secref{sec:entity}). After obtaining the entities, \toolname extracts the relationships among different entities according to the pre-defined relationships (\secref{sec:relation}). It is an iterative process to add information to the EKG with a one-by-one analysis of the crowdsourced test reports. During the accumulation, the coreference resolution (\secref{sec:coreference}) part merges redundant information by linking different descriptions of the same entities, aiming to enrich the entities from different perspectives of human testers.

\begin{algorithm}[h]
	\caption{EKG Construction}
	\label{alg:construction}
	\begin{algorithmic}[1]
		\Require Crowdsourced test report set $\mathbbm{R}$
		\Ensure EKG $\mathbbm{G}$
		\For{$report \in \mathbbm{R}$}
			\State $Entity_{image} \leftarrow$ extractEntity($report.image$)
			\State $Entity_{text} \leftarrow$ extractEntity($report.text$)
			\State $Entity \leftarrow Entity_{image} + Entity_{text}$
			\For{$entity \in Entity$}
				\State $type \leftarrow$ identifyEntityType($entity$))
				\State $\mathbbm{G}$.appendEntity($entity, type$)
			\EndFor
			\For{$entity_a \in Entity$}
				\For{$entity_b \in Entity$}
					\State $relation \leftarrow$ relationRecog($entity_a, entity_b$)
					\State $\mathbbm{G}$.appendRelation($entity_a, entity_b, relation$)
				\EndFor
			\EndFor
			\For{$entity_a \in Entity$}
				\For{$entity_b \in Entity$}
					\If{corefResol($entity_a, entity_b$) == True}
						\State addSimilarRelation($entity_a, entity_b$)
						\State $\mathbbm{G}$.update()
					\EndIf
				\EndFor
			\EndFor
		\EndFor
		\State return $\mathbbm{G}$
	\end{algorithmic}
\end{algorithm}

\subsection{Entity Extraction}
\label{sec:entity}

The entity extraction is conducted on the raw crowdsourced test reports. A crowdsourced test report consists of three parts, app screenshots, textual descriptions, and environment information. We use the app screenshots and textual descriptions for entity extraction. For the crowdsourced test reports, textual descriptions are composed of reproduction steps and bug descriptions, which present the steps that trigger the bugs from the app launch to bug occurrence and the app behavior when the bug occurs attached with expectations to correct behaviors, respectively. Each reproduction step can be matched to an app screenshot. However, it is possible that some test reports may lack some steps in textual descriptions or app screenshots. In order to alleviate the potential negative effects, we use different test reports to complementarily construct the EKGs to avoid potential information lack in some test reports. The complementation is reflected in two aspects. First, the information from different reports can help build a more complete entity similarity relationship. Different crowdworkers may use different descriptions to the same targets, which may fit in the corresponding scenarios in different apps. This can help better match the entities in the EKG and the widgets from the apps under test. Second, the workflows of different crowdworkers are different. Combining different test reports can help better explore different sub-scenarios that can complete the target scenarios by combining the different scenario-based testing paths. When faced with possible lack of information in some reports, like missing steps or entities, such information missing can be complemented from other reports by the relationship recognition with other entities or steps. Examples are provided in \secref{sec:example} and \figref{fig:exreport}.

App screenshots contain static information about the app states, and textual descriptions contain dynamic information about the testing operations. Such information can be extracted as entities to guide scenario-based mobile testing.

Textual descriptions consist of steps of test events, and can be abstracted as a list: $TE = \{op_1, op_2, ..., op_n\}$, where $op_i = \langle operation, widget, parameter \rangle$. The $operation$ refers to specific operations, like \texttt{click}, \texttt{input}, the $widget$ refers to the target of operations, and the $parameter$ refers to the attached information of the operation, like the text string of the input operation. The textual descriptions in natural language are processed with NLP technologies. Specifically, we use the jieba library\footnote{\url{https://github.com/fxsjy/jieba}} to first segment the sentences into words, and remove the unnecessary stop words. Jieba is a famous library to process the sentences into words for its effectiveness and efficiency. Then, we use an open-sourced dependency parsing analysis tool, DDParser, from Baidu\footnote{\url{https://GitHub.com/baidu/DDParser}} \cite{zhang2020practical} to reveal the syntactic structure of the sentence. Dependency parsing analysis \cite{zhu2013fast} is used to analyze the dependency relationship of the words, and then to determine the sentence structure. For textual descriptions in crowdsourced test reports in our scenario, there are several widely used sentence structures that can be utilized in the EKG generation: SBV (Subject-Verb structure), VOB (Verb-Object structure), CMP (Complement structure), ADV (Adverbial structure), and ATT (Attributive structure). The $operation$, $widget$, and $parameter$ correspond to the verb, object, and complement parts, respectively. Such information, including test operations, corresponding operation targets, necessary parameters, \etc, corresponds to the expected entity information for the EKG construction.

For the app screenshots, \toolname conducts processing of two aspects, the text aspect, and the non-text aspect. First, \toolname adopts OCR algorithms to extract all the existing texts from the app GUI screenshots. The coordinates of all the text fragments are recorded for the relationship recognition in \secref{sec:relation}. Second, with regard to the non-text aspect, \toolname uses the edge detection algorithms, \ie Canny, to extract the widget contours from the app GUI screenshots, and uses the morphological operations, including dilation and erosion, to characterize and extract the widgets, together with the widget coordinate information. On top of the widget images, \toolname uses a CNN model \cite{yu2021prioritize} to identify the widget type (\eg \texttt{Button}, \texttt{TextView}), to better help the relationship recognition and further EKG query in the scenario-based testing. Canny is a classic and efficient approach to detect edges from mobile app screenshots. Though it has been proposed for several years, many new studies are using it, like \cite{yu2021layout} \cite{yu2021prioritize} \cite{zhang2022unirltest}, and it is still effective and efficient. Besides, extracting widgets with Canny is only one step of \toolname, and we further use a CNN model to analyze the widget attributes to enhance the widget extraction based on the results of Canny. Therefore, after weighing the effectiveness and efficiency, we use the Canny algorithm to extract widgets from app screenshots.

During scenario-based mobile app testing, four kinds of entities are involved in the domain knowledge of human testers:

\noindent \textbf{\underline{Content (CNT)}} refers to the logic entities involved in the mobile app testing, and always refers to the concrete concepts. For example, the \texttt{Submit Button} is a CNT entity because it refers to a specific button on the app GUI.

\noindent \textbf{\underline{Widget (WID)}} refers to the widget types for all widgets obtained from the app GUI image, like \texttt{Button}, \texttt{TextField} \etc A WID entity does not refer to a specific widget on the app GUI, and it refers to an abstract widget concept.

\noindent \textbf{\underline{Operation (OPT)}} refers to specific operations, like \texttt{click}, \texttt{input}, \texttt{slide}, \etc The necessary parameters (like the input string, the sliding scale) are also attached.

\noindent \textbf{\underline{Text (TXT)}} refers to the texts extracted from the app GUI information, and some texts are linked to specific widgets.

For \toolname, we use the event knowledge graph \cite{chen2020review} instead of traditional knowledge graph technique, which is more suitable for a static situation. The reason is that scenario-based mobile app testing is a dynamic procedure, we need to indicate the order relationship among different entities, and some entities indicate the start or the end of the testing procedure. This is especially important for the Operation entities, which elaborate the actions that gradually push forward the testing procedures. Therefore, some Content entities are attached with two kinds of tags: start tag and tail tag. Start tags indicate whether a Content entity is a starting point of a testing scenario, and the tail tags indicate whether a Content entity is an ending point of a testing scenario. Such tags correspond to the first and last reproduction steps from the textual descriptions in crowdsourced test reports, which are in order according to the user operations. Besides, considering the sub-scenarios, the entities that may trigger different sub-scenarios will be labeled as the sub-scenario branch point.

\subsection{Relationship Recognition}
\label{sec:relation}

In the EKG of \toolname, logic CNT entities can store the test events and corresponding operation objects. Therefore, the other three kinds of entities will have relationships with the logic CNT entity. Based on the proposed entities, we define five different relationships as follows:

\noindent \textbf{\underline{TXT-TXT}}: \textit{similar} relationship is common in texts, including texts extracted from app screenshots and textual descriptions. Also, synonyms are considered in order to fit with the real-world testing process, \eg ``click'' or ``press'' on a button. The TXT-TXT relationship contains three situations: 1) texts extracted from app GUI are matched with the processed textual descriptions to link the widget images to the textual descriptions; 2) among texts of different crowdsourced test reports, \toolname calculates the text similarity to identify whether they describe the same logic CNT entity; 3) for widgets with texts, \toolname calculates the text similarity, which reflects the similarity of widget images. The \textit{similar} relationship is identified by the coreference resolution part (\secref{sec:coreference}).

\noindent \textbf{\underline{CNT-OPT}}: \textit{operate} relationship refers to the concept that operations can be applied on logic CNT entities, \eg ``click'' and ``button''. The CNT-OPT relationship involves the logic CNT entities and the OPT entities. The OPT entities are extracted from the textual descriptions, and the logic CNT entities in this relationship are always the abstract widget types, and every widget has one or more corresponding operations, \eg click on \texttt{Button} widgets.

\begin{figure}[!htbp]
\centering
\includegraphics[width=\linewidth]{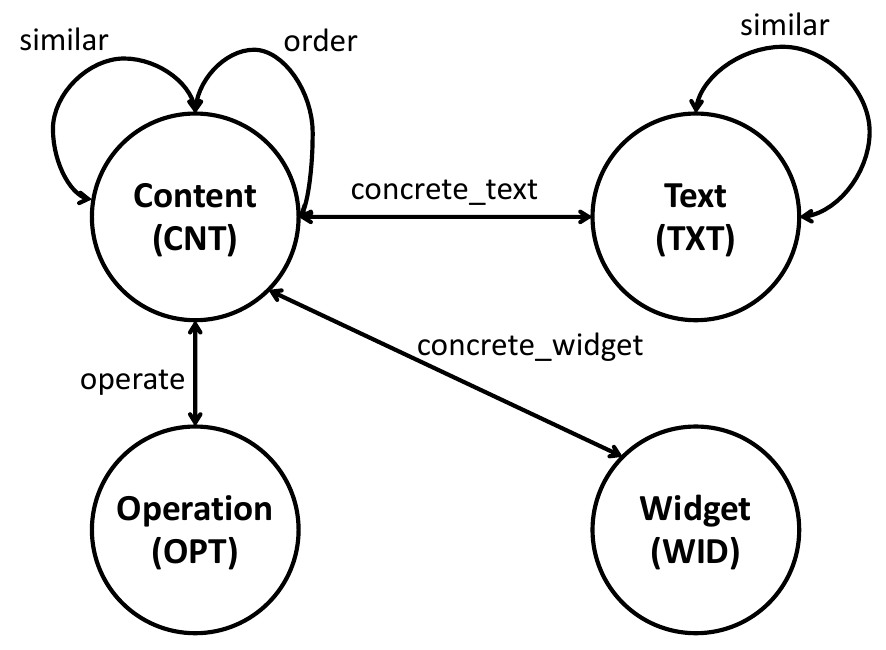}
\caption{Entities and relationships defined in the EKG for scenario-based mobile app testing.}
\label{fig:kg}
\end{figure}

\noindent \textbf{\underline{CNT-TXT}}: \textit{concrete\_text} relationship means logic CNT entities may be attached with some texts, \eg ``submit'' word on a ``button''. The CNT-TXT relationship links the concrete texts to the logic CNT entities. The texts are extracted from app GUI screenshots or textual descriptions and will enrich the attributes of the logic CNT entities, making it more clear during the EKG query.

\noindent \textbf{\underline{CNT-WID}}: \textit{concrete\_widget} relationship is among the logic CNT entities and abstract widget types \eg the ``Button'' type and the specific button on a screenshot. For all the entities and corresponding attribute information, two kinds of information models are involved: images and texts. For each image representing the GUI widgets, \toolname extracts the existing texts and obtains the widget type. Then the images can be linked to the logic CNT entities to form the CNT-WID relationship.

\noindent \textbf{\underline{CNT-CNT}}: \textit{similar} relationship means the similar logic CNT entities, which is used for the further coreference resolution and EKG query; \textit{order} relationship means the test event orders from crowdsourced test reports. The CNT-CNT relationship contains two sub-types: the similar relationship and the order relationship:

\begin{enumerate}[i]
	\item The \textit{similar} relationship measures the similarity among different logic CNT entities. For example, the ``\texttt{TextField}'' entity and the ``\texttt{TextBox}'' entity are actually similar logic CNT entities. The \textit{similar} relationship is identified by the coreference resolution part (\secref{sec:coreference}).
	\item The \textit{order} relationship indicates that for two specific logic CNT entities, there may exist the sequence order, which means that during the automated testing process, some logic CNT entities must be operated ahead of other ones. The order relationship also corresponds to the start tag and the tail tag of the logic CNT entities. The order relationship is identified from the reproduction steps from textual descriptions, which are in order according to the user operations.
\end{enumerate}

\subsection{Coreference Resolution}
\label{sec:coreference} 

Entities extracted from different crowdsourced test reports may be redundant due to the distinct language habits of various crowdsourced testing participants. Coreference resolution is necessary to identify the redundancy when constructing the event knowledge graphs iteratively with different crowdsourced test reports and to further merge the redundant information to complement the inadequate information of the target testing scenario.

Coreference resolution is closely related to the \textit{similar} relationship among entities. In the EKG design of \toolname, two kinds of entities have the \textit{similar} relationship: the Content and the Text. The redundant entities are not simply removed because different descriptions of the same concepts or logic entities will complement the information of different apps during the EKG-guided scenario-based testing of the testing scenarios of different mobile apps.

Specifically, coreference resolution is based on a two-step similarity calculation, and it can help form the \textit{similar} relationship, which is important in identifying the distinct descriptions of the same entities from different perspectives. \toolname involves similarity calculation among GUI images and texts. GUI images are attached with texts, so the attached texts can represent the GUI images. During the text similarity calculation, \toolname first uses a synonym dataset from \cite{yu2021prioritize} to match the keywords, considering the special meanings of some words under the software testing context. If no synonyms are matched, \toolname calculates the text similarity with the widely-used Euclidean distance after transforming the texts into vectors with the Word2Vec model. The coreference resolution part labels such redundant entities based on the similarity calculation and constructs the event knowledge graphs for testing scenarios based on the pre-defined entities and relationships.

\section{Scenario-based Mobile App Testing}
\label{sec:test}

Different from traditional automated testing approaches, which explore the AUT following specific strategies (random-based, model-based, or deep/reinforcement learning-based) to improve code coverage, scenario-based mobile app testing is guided by EKGs that are constructed based on the domain knowledge of human testers (as shown in the bottom half of \figref{fig:framework} and \agmref{alg:generation}). \toolname realizes the automated simulation of the human testing process, from the perspective of app GUI and with the consideration of app business logic.

For a specific mobile app, \toolname first conducts the semantic understanding of the app GUI of the current state (\secref{sec:gui}), including widget extraction, widget attribute obtaining, and GUI layout characterization. Then, such information will be restructured to a nested app GUI structure file and be sent to the EKG together with the testing context (For the first step, the testing context is null). The testing context is maintained during the testing process. When the information is passed to the EKG by a query, \toolname matches the existing widgets from the queries and entities in the EKG. The matched widgets on app current state will be assigned a probability for each according to the similarity. Such probabilities indicate which widget can push forward the scenario-based testing if it is operated. The query results, including the widgets and probabilities, are then returned. The operation on the widget with the highest probability will be operated. Besides, the returned widgets may be labeled to indicate whether they will start different sub-scenario branches or lead to an end of scenario-based testing.

The actually executed widgets will be added to the testing context. If the ending indicator (details in \secref{sec:case}) appears, the testing process of the current sub-scenario will be terminated. Then, \toolname will check whether there are any unexplored sub-scenario branches. If so, the app GUI will be redirected to the point and the exploration to a different sub-scenario will start with the initialization of the AUT. When all the sub-scenarios are explored, the whole process of scenario-based testing is considered complete.

\begin{algorithm}[h]
	\caption{Scenario-based Mobile App Testing}
	\label{alg:generation}
	\begin{algorithmic}[1]
		\Require App Current State $\mathbbm{S}$, EKG $\mathbbm{G}$
		\Ensure Test Event $\mathbbm{E}$
		\State initialize test context $\mathbbm{C}$
		\State $Entity_{app} \leftarrow$ widgetExtraction($\mathbbm{S}$)
		\For{$entity \in Entity_{app}$}
			\State $entity$.attach($entity$.attribute())
		\EndFor
		\State layoutCharacterization($\mathbbm{S}$)
		\For{$entity_{app} \in Entity_{app}$}
			\For{$entity_{EKG} \in \mathbbm{G}$}
				\If{match($entity_{app}, entity_{EKG}$)}
					\If{$entity_{app}$.match($\mathbbm{C}$)}
						\State $\mathbbm{E}$.add($entity_{app}$)
					\EndIf
				\EndIf			
			\EndFor
		\EndFor
		\State return $\mathbbm{E}$
	\end{algorithmic}
\end{algorithm}

\subsection{App GUI Analysis}
\label{sec:gui}

App GUI analysis targets at the app current state. \toolname has a semantic understanding: widget extraction, widget attribute obtaining, and GUI layout characterization.

\textbf{Widget Extraction.} The widget extraction is based on an edge detection algorithm, \ie Canny. Further, the morphological operations, including dilation and erosion, are applied to make widget contours more clear and to eliminate subtle elements that may be mistakenly recognized as widgets, making the widget recognition result more precise. We do not use the deep learning-based widget detection technologies \cite{chen2020object} to make the \toolname more lightweight.

\textbf{Widget Attribute Obtaining.} Besides the widget images, the widgets have more semantic information to explore. First, the widget coordinates are important, which can reflect the relationship among all the widgets. Second, the widget type is also significant, which is recognized with a CNN model \cite{yu2021prioritize}. The widget type can be used to be matched with specific operations, \eg \texttt{click} or \texttt{long-click} to a \texttt{Button}, \texttt{input} to a \texttt{TextField}. Third, the texts on the widgets are recognized with the OCR algorithm. We do not extract texts from the GUI layout for the following reasons. First, the GUI layout may lose information in some special widgets \cite{yu2021layout}, like the ``canvas'' widget, where inner contents are not available. Some widget information from the embedded HTML pages is also not accessible. Such information will be definitely presented on the app GUI for the users. Therefore, we think app GUI screenshot is a better source to extract complete text information. Second, existing OCR algorithm has achieved huge advancement and can recognize characters effectively, even for multilingual situation. Therefore, we believe it is an appropriate choice to use OCR to extract texts instead of extracting from the GUI layout. Then, the semantic attributes of widgets are bound to the corresponding widgets, making it easier to match with the entities in the EKG.

\textbf{GUI Layout Characterization.} The extracted widgets, together with attributes, are sent as a query to the EKG for entity matching. Layout characterization is based on the coordinates of the widgets. The vertical coordinates are first used to form the horizontal layout hierarchy, and the horizontal coordinates are then used to form the vertical layout hierarchy. The GUI widgets are organized into the tree structure, and the attributes are attached to the tree nodes (\ie widgets). For the widgets that have merely slight differences with regard to the vertical or horizontal coordinates, we set a threshold (default as 0.1 of the app screenshot size according to \cite{yu2021layout}) to merge the close coordinates and reduce unnecessary layout hierarchies.

\begin{table*}[!htbp]
\centering
\caption{Event Knowledge Graph Ontology}
\scalebox{0.9}{

\begin{tabular}{c|r|r|r|r|r|r|r|r|r|r|r}
\toprule
Scenario     & \multicolumn{1}{c|}{\textbf{Entity}} & \multicolumn{1}{c|}{Content} & \multicolumn{1}{c|}{Widget} & \multicolumn{1}{c|}{Operation} & \multicolumn{1}{c|}{Text} & \multicolumn{1}{c|}{\textbf{Relationship}} & \multicolumn{1}{c|}{TXT-TXT} & \multicolumn{1}{c|}{CNT-OPT} & \multicolumn{1}{c|}{CNT-TXT} & \multicolumn{1}{c|}{CNT-WID} & \multicolumn{1}{c}{CNT-CNT} \\ \midrule

Login        & \textbf{300}  & 127 & 17  & 7  & 149 & \textbf{2481} & 722  & 190 & 595  & 470 & 504  \\
Register     & \textbf{231}  & 88  & 17  & 5  & 121 & \textbf{1002} & 150  & 88  & 157  & 163 & 444  \\
Email        & \textbf{73}   & 13  & 17  & 3  & 40  & \textbf{188}  & 10   & 23  & 75   & 44  & 36   \\
FlightTicket & \textbf{73}   & 16  & 17  & 4  & 36  & \textbf{233}  & 86   & 22  & 50   & 37  & 38   \\
AddCart   & \textbf{82}   & 15  & 17  & 3  & 47  & \textbf{206}  & 42   & 15  & 57   & 50  & 42   \\ 
Chat         & \textbf{90}   & 18  & 17  & 4  & 51  & \textbf{221}  & 51   & 19  & 60   & 40  & 51   \\ 
Music        & \textbf{89}   & 21  & 17  & 5  & 46  & \textbf{226}  & 54   & 21  & 62   & 36  & 53   \\ 
Video        & \textbf{98}   & 23  & 17  & 3  & 55. & \textbf{192}  & 28   & 31  & 57   & 37  & 39   \\ \midrule
Sum          & \textbf{1036} & 321 & 136 & 34 & 545 & \textbf{4749} & 1143 & 409 & 1113 & 877 & 1207 \\ \bottomrule
\end{tabular}

}

\label{tab:kg}
\end{table*}

\subsection{Scenario-based Test Generation}
\label{sec:case}

During the scenario-based test generation, the guidance from the EKG is mainly based on two aspects of information, the current app state (GUI information), and the testing context. Mobile app testing is an event-driven process \cite{su2017guided}, so the context can greatly affect the operation selection.

\textbf{\textit{\underline{Sub-Scenario}}}. Intuitively, the activity transition of mobile apps can be seen as a directed graph. The scenario can be viewed as going from one graph node (app activity) to another, and there may exist several paths that can satisfy the goal. That is to say, in order to complete the scenario-based testing, different execution paths can achieve the same testing goals. Therefore, we define the Sub-Scenario concept, which means that one independent path can complete the testing scenario. During scenario-based testing, the test events of all sub-scenarios are supposed to be generated. For example, for the login scenario, users can log in through the combination of username and password, and they can also log in with a phone number and verification code. Sub-scenario is more like a skeleton that guides the execution of test event sequences. In our design of \toolname, a sub-scenario represents an execution path that can complete the testing scenario. One sub-scenario may take different inputs (valid or invalid) and may result in different testing results, while different inputs will go through the same execution path. Valid inputs are supposed to successfully complete the sub-scenarios but invalid inputs may lead to the failure of the sub-scenarios. The failure situation is also considered in the design of \toolname. Take the Login scenario as an example, a correct username-password pair can lead to a successful login result, but a mismatched pair may lead to a half-way termination of the test generation. In the current design of \toolname, we take the inputs identified in the crowdsourced reports or randomly generated inputs to push forward the process.

To obtain guidance from the EKG to generate test events for the scenarios, the app current state and the context will be sent to the EKG. The app currently is analyzed into a nested app GUI structure file as described in \secref{sec:gui}, and the testing context is maintained as a sequence of test events. The whole test generation process can be viewed as a depth-first exploration, which means that when one sub-scenario is done, the app state will be recalled to the branch point and starts the next sub-scenario until all the sub-scenarios are finished, which indicates the accomplishment of the scenario-based testing exploration.

When the app current state is analyzed and formed as a query to the EKG, all the widget elements in the nested app GUI structure file will be matched with the entities of the EKG. In order to find one GUI widget to operate on the app GUI that mostly fits the EKG entities in each query we consider two aspects: whether the widgets on the app current GUI state are in the proper sub-scenario path (context) according to the query to the EKG, and the similarities between the widgets on the app current GUI state and the entities in the EKG. In order to determine the widget to operate, we adopt a two-step process to calculate the execution probability $P$, which is defined in \equref{equ:prob}.

\begin{equation}
\label{equ:prob}
	P = \left\{ 
	\begin{aligned}
		& 0, & w \notin context \\ 
		& similarity(w, e), & w \in context
	\end{aligned}
	\right.
\end{equation}

where $w$ refers to each widget extracted from the app GUI, $e$ refers to each entity in the EKG, and $context$ refers to the proper sub-scenario path in the EKGs.

For the first step, \toolname checks whether the matched widgets are in the proper sub-scenario path. In other words, the nodes with \textit{order} relationships linked to app current GUI state node in EKG are matched to the exploration context, which shows the explored app GUI states. If the nodes in EKG and the exploration context are matched, the matched widget is supposed to be in the proper sub-scenario path. If the widgets are not in the proper sub-scenario path, which means the widgets are in other parts of the EKG other than the target sub-scenario path, the probability will be directly set as 0. One thing to notice is that in the first event, the widgets will only be matched with entities with the starting point label. For those widgets are in the proper sub-scenario path (\ie context), we calculate the similarity in the second step. For the second step, the widgets in the app current state will find the closest entity according to the similarity calculation illustrated in \secref{sec:coreference}. The obtained similarity is considered the probability of the widgets being operated.

Finally, \toolname chooses to operate on the widget with the highest probability, and only one widget will be actually operated after the probability calculation. The widget with the highest probability will be added to the testing context after being operated. Specifically, the input content for \texttt{input} operation is from the seed samples. The returning widgets are likely to be labeled with tail tags. If the widget to be operated is with the tail tag, it indicates the completion of one sub-scenario. 

When one sub-scenario is finished, the app state will be recalled to the closest state that indicates a branch point of different sub-scenarios. Such branch points are also recorded during the EKG construction. Then the exploration of new sub-scenarios starts. During the app exploration, execution of different sub-scenarios may lead to potential interference. We design the app state reinitialization to avoid the interference. We record the test events in context during the app exploration. When one sub-scenario is completed and another is to start, the app is restarted and the test events from app launch to the new sub-scenario branch will be executed. When all the sub-scenarios are explored (judged by the recorded branch points), the scenario-based testing on the AUT is finished.

The tail tag indicates the normal end of the exploration. \toolname combines with the query results to identify the completeness of exploration to sub-scenarios:

\begin{itemize}
	\item The last test event is labeled with a tail tag, and the result set of the EKG query is null: the sub-scenario is finished.
	\item The last test event is not labeled with a tail tag, and the result set of the EKG query is not null: the current sub-scenario is not completed, and further test generation is possible.
	\item The last test event is not labeled with a tail tag, and the result set of the EKG query is null: the test generation conflicts with the EKG and scenario-based testing meets a failure. The testing on the current sub-scenario will be terminated and the testing on the next sub-scenario will start by reinitializing the app state to the new sub-scenario branch point.
\end{itemize} 

\section{Experiment}
\label{sec:exp}

\subsection{Experimental Settings}

We set five research questions (RQ) from different aspects to completely evaluate \toolname. The first two RQs are targeted at the EKG construction. RQ1 evaluates how correct the event knowledge graph is with the automated technologies of \toolname. RQ2 evaluates the reliability of the constructed EKGs. The RQ3 and the RQ4, targeted at the scenario-based mobile app testing effectiveness, evaluate the scenario-based testing with the guidance of EKG from the accuracy and adequacy, respectively. RQ5 evaluates the usefulness of \toolname in finding real-world bugs in specific testing scenarios.

\begin{itemize}
	\item \textbf{RQ1 (Correctness)}: Can \toolname construct correct event knowledge graphs for scenario-based testing?
	\item \textbf{RQ2 (Reliability)}: Can the EKG of \toolname provide reliable predictions for scenario-based testing?
	\item \textbf{RQ3 (Accuracy)}: How accurately can \toolname test all the scenarios on different apps based on the EKG?
	\item \textbf{RQ4 (Adequacy)}: Can \toolname cover adequate scenario branches on different apps?
	\item \textbf{RQ5 (Usefulness)}: Can \toolname effectively reveal bugs for specific scenarios?
\end{itemize}

\begin{table}[!htbp]
\centering
\caption{Apps in the Empirical Evaluation}
\scalebox{1}{
\begin{tabular}{c|r|r|r|r}
\toprule
  Scenario     
& \multicolumn{1}{c|}{\thead{EKG\\Construction}} 
& \multicolumn{1}{c|}{Testing} 
& \multicolumn{1}{c|}{\thead{\# Sub-\\Scenario}} 
& \multicolumn{1}{c}{Event} 
\\ \midrule
Login        & 10 & 30 & 37 & 245 \\
Register     & 10 & 30 & 59 & 343 \\
Email        & 7  & 17 & 10 & 71  \\
FlightTicket & 8  & 19 & 11 & 49  \\
AddCart      & 10 & 21 & 10 & 53  \\ 

Chat         & 7  & 16 & 15 & 47  \\ 
Music        & 8  & 19 & 20 & 66  \\ 
Video        & 10 & 21 & 23 & 71  \\ \midrule

\thead{Sum\\(with repetition)} & 70 & 124 & 185 & 945 \\ \bottomrule
\end{tabular}}
\label{tab:app}
\end{table}

\textit{\textbf{Experimental Subjects}} In this paper, we in total focus on eight common testing scenarios to evaluate the proposed \toolname, including Login, Register, Email, FlightTicket, AddCart, Chat, Music, and Video. During the selection of the subject scenarios, we follow two criteria:

\begin{enumerate}[i]
	\item The scenario should be widely available in apps of different categories.
	\item The scenario should be widely available in one or more specific categories of apps that are quite common and popular.
\end{enumerate}

The selected scenario should at least satisfy one of the criteria. For the first criterion, we select the Login and Register scenarios, which are no doubt the most common scenarios in all categories of apps. For the second criterion, we both refer to the Google app store and a widely used mobile app research benchmark, AndroZooOpen \cite{liu2020androzooopen}, which respectively reflect the popularity of commercial and open-source mobile apps. From the Google app store and AndroZooOpen dataset, we find the most popular app categories according to the app quantities and the download/star numbers. Then we obtain the email, travel, shopping, social, and media apps. Among these app categories, we use their most common and primary business scenarios, which are the Email, FlightTicket, AddCart, Chat, Music, and Video scenarios.

Login refers to using a specific method to have access to some services, like using the combination of username and password or the combination of phone number and verification code. Register refers to creating a new account with some personal information, like using only username and password or sometimes more extra information. Email refers to sending an email, like directly starting a new email or replying to an existing email. FlightTicket refers to booking a flight ticket with a search, like by directly searching a flight or by searching the destination. AddCart refers to searching for a good and adding to the shopping cart, like searching by fuzzy matching or condition filtering. Chat refers to sending messages, including text, image, \etc Music refers to finding a music by searching title or other attributes, and playing it. Video refers to finding a target video by searching title or other attributes, and playing it. Our evaluation covers simple and complex scenarios. For simple scenarios, like Login and Register, the sub-scenarios are usually more complex, and are more common in different apps. For complex scenarios, like Email and AddCart, the sub-scenarios are usually less complex, and may only appear in specific categories of apps. No matter the scenarios are common in different app categories or only appear in specific categories, \toolname can perform well. Therefore, we believe the scope of our experiment subjects is sound by considering large number of apps and scenarios of different complexities, and \toolname is supposed to adapt more scenarios in different apps. Our preliminary investigation of the app stores shows that these testing scenarios are most common in different categories of mobile apps. Login and Register scenarios are common in all different apps. Email, FlightTicket, AddCart, Chat, Music, and Video scenarios are common in apps of specific categories, which are the most common categories in app stores and have an important impact on people's daily life. 

In total, we use a dataset containing 22,720 crowdsourced test reports from 70 real-world mobile apps with large popularity to construct the event knowledge graphs \cite{yu2021prioritize}. The crowdsourced test reports are collected on such apps from a widely-used crowdsourced testing platform, MoocTest\footnote{\url{http://www.mooctest.net}}, one of the representative crowdsourced testing platforms in China. MoocTest has supported many academic studies in crowdsourced test report analysis \cite{yu2021prioritize}. There are over 40,000 users from over 1,000 affiliations, spanning among over 50 countries (data until Dec 2023), which shows the great popularity and representativeness of the platform. Besides, MoocTest adopts almost identical crowdsourced testing mechanism with large-scale commercial platforms, like Global App Testing, Baidu Crowdtesting, Testin, \etc Therefore, we believe that the reports from Mooctest can show the generalizability of the EKG construction source and method of \toolname. Another 124 apps are used for the effectiveness evaluation in RQ1 to RQ4. For RQ5, we use a subset of 124 apps (numbers shown in \tabref{tab:rq5}), which are all open-sourced apps, because the apps needed to be instrumented to collect the testing logs and to capture bugs. The involved apps, like \textit{BBC}, \textit{CNN},\textit{Wikipedia}, \textit{Amazon}, \textit{Taobao}, \textit{Spotify}, \textit{Outlook}, \textit{Gmail}, \textit{Lufthansa}, \textit{eDreams},\textit{eSky}, \etc, cover many different categories, including Business, Entertainment, Tool, Communication, Education, \etc, which shows the generalizability of \toolname. We use apps from previous studies, including \cite{su2017guided} and \cite{liu2020androzooopen}. One thing to notice is that one app may be used for more than one testing scenario, but none of the apps are used for both EKG construction and scenario-based testing. We also reveal more details in our online package. \tabref{tab:app} shows the app quantities for EKG construction and scenario-based testing, respectively. We believe the apps and scenarios we use in the evaluation are with enough representativeness. The Login and Register scenario is relatively simple but the sub-scenarios are more diverse. The rest scenarios are complex in business logic. 

The EKG ontology is shown in \tabref{tab:kg}, which shows the total entity number and relationship number for eight testing scenarios, and the respective numbers of four entities and five relationships. The event knowledge graphs for eight testing scenarios involve 1036 entities and 4749 relationships.

\subsection{RQ1: Correctness}

First, we evaluate whether \toolname can effectively identify the operations and corresponding objects. The four kinds of entities are in the form of images from app screenshots and texts from both textual descriptions and app screenshots, so the evaluation target is the images and texts. We analyze the 945 test events in the event knowledge graph, and the test events refer to the operations on specific widgets. We invite three third-party testing experts to manually evaluate the extraction results. The experts are with at least five years of experience in Android development and testing, and they are familiar with different apps with the target testing scenarios of the experiment. During the manual evaluation, the experts are required to identify all the operations and corresponding objects, and different experts cross-validate the results to reach a consensus. During the evaluation, they will have discussions to resolve potential disagreements, and the Cohen’s Kappa value is 0.93, which shows the almost perfect agreement. The results are shown in the first two columns of \tabref{tab:rq1}. We can see that the overall extraction success rate is 99.05\% ($936 / 945*100\%$), which shows the excellent performance of the entity extraction from texts. Furthermore, the average precision and recall value of the eight testing scenarios reach 96.98\% and 99.02\%, indicating that \toolname not only can effectively extract entities, but also has very few mis-recognized entities during the entity extraction. We then investigate the failures and conclude two main causes: the wrong word segmentation, and the omitting adjectives. For example, one report says ``click on the green button'', while the extracted object is only ``button'' (an occasional sample), and there are several buttons on the screenshots, so the result may lead to misunderstandings, which would be considered a failure.

\begin{table}[!h]
\caption{Experiment Results for RQ1: Correctness}
\centering
\scalebox{0.77}{
\begin{tabular}{c|rrrr|rrr}

\toprule
\multicolumn{1}{c|}{Scenario}
& \multicolumn{1}{c}{\thead{Test\\Event}}   
& \multicolumn{1}{c}{\# Succ}        
& \multicolumn{1}{c}{Precision}  
& \multicolumn{1}{c|}{Recall}    
& \multicolumn{1}{c}{\thead{Text\\Match}} 
& \multicolumn{1}{c}{\thead{Layout\\Match}} 
& \multicolumn{1}{c}{\% Succ}                                               
\\  \midrule

Login        & 245 & 242 & 99.18\%  & 98.78\%  & 204 & 29  & 95.10\%  \\
Register     & 343 & 340 & 99.71\%  & 99.13\%  & 287 & 52  & 98.83\%  \\
Email        & 71  & 71  & 94.67\%  & 100.00\% & 39  & 32  & 100.00\% \\
FlightTicket & 49  & 49  & 94.23\%  & 100.00\% & 17  & 28  & 91.84\%  \\
AddCart      & 53  & 53  & 98.15\%  & 100.00\% & 21  & 31  & 98.11\%  \\

Chat         & 47  & 45  & 95.74\%  & 95.74\%  & 28  & 17  & 95.74\%  \\ 
Music       & 66  & 65  & 95.59\%  & 98.48\%  & 34  & 29  & 95.45\%  \\ 
Video        & 71  & 71  & 98.61\%  & 100.00\% & 38  & 30. & 95.77\%  \\ \midrule

Sum / Avg    & 945 & 936 & 96.98\%  & 99.02\%  & 668 & 248 & 96.36\%  \\ \bottomrule

\end{tabular}}
\label{tab:rq1}
\end{table}

Second, we evaluate the text-image matching effectiveness, which is important to relationship recognition and coreference resolution. We use manual evaluation in this part. The results are shown in the last three columns of \tabref{tab:rq1}. We can see that for the eight testing scenarios, the average success rate reaches 96.36\%, which is a good performance. During the EKG construction, \toolname can effectively match the widget images to the extracted texts. Then, we investigate the failed cases and conclude three main causes: wrong text extraction, mistake reports, and non-text widget reference. The first cause is due to mistakes caused in the text extraction, which leads to chain reactions. The second cause is due to the mistakes in the reports not matching app GUI screenshots. The third cause is due to the widgets that are hard to be identified. A typical example is the image-based non-robot verification during the Login testing scenario. 

In summary, \toolname can effectively extract entities from the crowdsourced test reports, and it can effectively match the images extracted from app screenshots and texts from textual descriptions in crowdsourced test reports. 

\subsection{RQ2: Reliability}

In order to evaluate the reliability of the prediction results of the EKG, we record all the GUI screenshots on the apps for scenario-based testing according to the sub-scenarios recorded in the EKG manually, and we also label all the widgets on the screenshots whether they are expected to be operated (ground-truth). For each GUI screenshot, we take it as the app current state to feed into the EKG, and the first prediction result of the EKG querying result list is then matched with the manual labeling ground-truth. 

\begin{table}[!h]
\caption{Experiment Results for RQ2: Reliability}
\centering
\scalebox{0.85}{
\begin{tabular}{c|rrr}

\toprule
\multicolumn{1}{c|}{Scenario}
& \multicolumn{1}{c}{Test Event}   
& \multicolumn{1}{c}{\# Succ}
& \multicolumn{1}{c}{\% Succ}                                               
\\  \midrule

Login        & 245 & 242 & 98.78\% \\
Register     & 343 & 318 & 92.71\% \\
Email        & 71  & 67  & 94.37\% \\
FlightTicket & 49  & 46  & 93.88\% \\
AddCart      & 53  & 50  & 94.34\% \\
Chat         & 47  & 44  & 93.62\% \\
Music        & 66  & 63  & 95.45\% \\
Video        & 71  & 67  & 94.37\% \\ \midrule

Sum / Avg    & 945 & 897 & 94.69\% \\ \bottomrule

\end{tabular}}
\label{tab:rq2}
\end{table}

From \tabref{tab:rq2}, we can see that the average success rate of predicting the widget to be operated is 94.69\%, ranging from 92.71\% to 98.78\% among different scenarios. The average success rate refers to the ratio of the number of successfully predicted results and the number of all predictions provided by EKG. The results show that \toolname can effectively identify the target widgets from the app current state and provide reliable prediction results.

\subsection{RQ3: Accuracy}

With the automatically constructed EKGs, it is important to evaluate the effectiveness of how accurate \toolname can generate tests and test specific scenarios. We set a metric $GenRate$ to evaluate whether \toolname can generate effective test cases, which is calculated as: 

\begin{equation}
	GenRate = \frac{\#~TestOp}{\#~ScenOp} \times 100\%
\end{equation}

where $TestOp$ refers to the actual generated operations, and the $ScenOp$ is the required operations from the EKG. In other words, $ScenOp$ refers to the operations in the EKG that can indispensably complete the target scenario. $TestOp$ and $ScenOp$ are the same form as the $op_i$ defined in \secref{sec:entity}. If the target widget, the expected operation, and the corresponding parameter of one $TestOp$ and one $ScenOp$ are exactly the same, we consider the test generation successful. When using the $GenRate$ metric, the sequence order is implied in the evaluation. If one test event is mistakenly generated, all its subsequent generated test events are considered failures. Such a design further illustrates the reasonability of the $GenRate$ design by considering the temporal influence of the generated tests. The results can be seen from \tabref{tab:rq3}, and the average $GenRate$ reaches 87.74\%, which is a high success rate. \toolname achieves the best result in the FlightTicket scenario, which is 92.31\%, and the performance on Login and Register scenarios is relatively low, which is close to 85\%. The reason is that these two scenarios are almost available on all different apps, and the scenarios are relatively complex. They may contain more customized designs regarding their own business logic, even if the overall workflow or general business logic is similar. The results show that the proposed \toolname is capable of successfully generating scenario-based test cases with the guidance of EKG for different mobile apps.

\begin{table}[!h]
\caption{Experiment Results for RQ3: Accuracy}
\centering
\scalebox{0.85}{
\begin{tabular}{c|rrr}

\toprule
\multicolumn{1}{c|}{Scenario}
& \multicolumn{1}{c}{\# ScenOp} 
& \multicolumn{1}{c}{\# TestOp} 
& \multicolumn{1}{c}{GenRate}                                              
\\  \midrule

Login        & 192 & 163 & 84.90\% \\
Register     & 161 & 137 & 85.09\% \\
Email        & 69  & 61  & 88.41\% \\
FlightTicket & 39  & 36  & 92.31\% \\
AddCart      & 85  & 78  & 91.76\% \\
Chat         & 78  & 69  & 88.46\% \\
Music        & 82  & 73  & 89.02\% \\
Video        & 72  & 63  & 87.50\% \\ \midrule

Sum / Avg    & 778 & 680 & 88.43\% \\ \bottomrule

\end{tabular}}
\label{tab:rq3}
\end{table}

Moreover, we have a deep investigation into the failures, which are mostly caused by random and uncontrollable events. First, system events may interrupt the scenario-based testing, like incoming calls. Second, apps may encounter image-based non-robot verifications that cannot be automatically processed. Third, some scenarios may involve outer app transition to ask for third-party app authorization, which may have difficulty in returning to the AUT. These uncontrollable events do not necessarily indicate bugs in the AUT.

\subsection{RQ4: Adequacy}

The adequacy is also an important indicator of the effectiveness of the scenario-based testing. In order to judge adequacy, coverage is a useful metric. Traditional automated approaches tend to use code coverage, while in our approach, we use the sub-scenario as the coverage object. Different sub-scenarios consist of the whole testing scenario, so covering all the sub-scenarios is critical.

In order to answer the RQ4, we introduce a metric called $ScenCov$, and is calculated as

\begin{equation}
	ScenCov = \frac{|S_t|}{|Scen|} \times 100\%
\end{equation}

where $S_t$ refers to the set of sub-scenarios that are automatically explored by the \toolname, and the $Scen$ refers to the set of all sub-scenarios of a specific testing scenario. 

\begin{table}[!h]
\caption{Experiment Results for RQ4: Adequacy}
\centering
\scalebox{0.85}{
\begin{tabular}{c|rrr}

\toprule
\multicolumn{1}{c|}{Scenario}
& \multicolumn{1}{c}{\# Sub-Scenario}
& \multicolumn{1}{c}{\# Covered Scenario}
& \multicolumn{1}{c}{ScenCov}                                            
\\  \midrule

Login        & 67  & 55  & 82.09\%  \\
Register     & 39  & 33  & 84.62\%  \\
Email        & 24  & 20  & 83.33\%  \\
FlightTicket & 8   & 8   & 100.00\% \\
AddCart      & 15  & 12  & 80.00\%  \\
Chat         & 31  & 27  & 87.10\%  \\
Music        & 29  & 26  & 89.66\%  \\
Video        & 36  & 33  & 91.67\%  \\ \midrule

Sum / Avg    & 249 & 214 & 87.31\%  \\ \bottomrule

\end{tabular}}
\label{tab:rq4}
\end{table}

The results are presented in \tabref{tab:rq4}. The average $ScenCov$ of eight testing scenarios reaches 87.31\%, and \toolname has the best performance on the FlightTicket scenario, which is 100\%. The results indicate that \toolname can effectively test adequate sub-scenarios. Adequacy means whether \toolname can effectively go through all the sub-scenarios under specific scenarios. Only if all the sub-scenarios are tested, the scenarios can be fully tested. The failures are investigated. Some sub-scenarios are app-specific, and cannot be adapted to other apps. Such sub-scenarios are quite rare in other apps even with the same testing scenario.

\subsection{RQ5: Usefulness}

\begin{table*}[!htbp]
\centering
\caption{Experiment Results for RQ5: Usefulness}
\scalebox{1}{
\begin{tabular}{cr|rrrrr|rrrrr}
\toprule
\multicolumn{1}{c}{Scenario}    
& \multicolumn{1}{c|}{\# App} 
& \multicolumn{1}{c}{M - Sc (all)}
& \multicolumn{1}{c}{M - Sc (sce)} 
& \multicolumn{1}{c}{M $\cap$ Sc} 
& \multicolumn{1}{c}{\textbf{Sc - M}} 
& \multicolumn{1}{c|}{Total} 
& \multicolumn{1}{c}{St - Sc (all)} 
& \multicolumn{1}{c}{St - Sc (sce)} 
& \multicolumn{1}{c}{St $\cap$ Sc}
& \multicolumn{1}{c}{\textbf{Sc - St}} 
& \multicolumn{1}{c}{Total}   \\ \midrule

Login        & 15 & 329  & 0 & 42  & \textbf{33}  & 75  & 343  & 0 & 42  & \textbf{33}  & 75  \\ 
Register     & 15 & 287  & 2 & 20  & \textbf{18}  & 40  & 301  & 2 & 19  & \textbf{18}  & 40  \\ 
Email        & 12 & 143  & 0 & 13  & \textbf{15}  & 28  & 152  & 0 & 11  & \textbf{17}  & 28  \\ 
FlightTicket & 7  & 142  & 0 & 3   & \textbf{9}   & 12  & 151  & 0 & 3   & \textbf{9}   & 12  \\ 
AddCart      & 9  & 128  & 0 & 12  & \textbf{10}  & 22  & 135  & 0 & 11  & \textbf{11}  & 22  \\ 
Chat         & 6  & 133  & 2 & 15  & \textbf{18}  & 35  & 143  & 2 & 14  & \textbf{20}  & 35  \\ 
Music        & 7  & 129  & 1 & 14  & \textbf{34}  & 49  & 144  & 1 & 12  & \textbf{36}  & 49  \\ 
Video        & 6  & 102  & 1 & 12  & \textbf{21}  & 34  & 124  & 0 & 6   & \textbf{28}  & 34  \\ \midrule
Sum          & 77 & 1393 & 6 & 131 & \textbf{158} & 295 & 1493 & 5 & 118 & \textbf{172} & 295 \\ \bottomrule

\end{tabular}}
\label{tab:rq5}
\end{table*}

In this research question, we hope to evaluate the usefulness of the \toolname, which means whether \toolname can effectively find bugs for specific scenarios. We use two representative and widely-used baselines: Monkey \cite{google2022monkey} and Stoat \cite{su2017guided}. Stoat is a representative and state-of-the-art tool for automated mobile app exploration testing. We run the baselines for two hours on each app and collect the bugs. For \toolname, there is no time limit because it will stop when the iteration is finished, and we collect the time overhead data. On average, \toolname can complete the scenario-based testing on an app in 378.29 seconds, ranging from 152 seconds to 993 seconds, depending on the EKG complexity, sub-scenario numbers, app GUI complexity, \etc 

In this paper, we focus on specific scenarios to compare the approaches. The results of the revealed bugs are shown in \tabref{tab:rq5}. One thing to notice is that the exploration scope of Monkey and Stoat is the whole app, so it can find more bugs beyond the target testing scenario. We list the detected bug numbers under the ``M - Sc (all)'' and ``St - Sc (all)'' columns for a reference. We manually analyze and cross-validate the logs of Stoat and Monkey based on the activity and source code package information. Then, the revealed bugs having no relationship to specific scenarios are filtered out. Then, we filter the duplicate bugs with a very careful manual check, which is based on the exception texts, code line, activity name, and source code file name in the testing logs. The bugs revealed are all crash bugs. In \tabref{tab:rq5}, the \textit{Monkey/Stoat - Sc (sce)} refers to the number of bugs revealed by Monkey/Stoat but not revealed by \toolname. \textit{Monkey/Stoat $\cap$ Sc} refers to the number of bugs revealed both by Monkey/Stoat and \toolname. \textit{Sc - Monkey/Stoat} refers to the number of bugs revealed by \toolname but not revealed by Monkey/Stoat. The results show that \toolname can find 158 and 172 distinct bugs compared with Monkey and Stoat, respectively. The number accounts for 53.56\% and 58.31\% of the total revealed bugs of specific testing scenarios. That is to say, half of the revealed bugs in the 77 apps of specific testing scenarios can be only revealed with the guidance of domain knowledge. In contrast, Monkey and Stoat can only find 6 and 5 distinct bugs (5 bugs revealed by Stoat are covered by Monkey). 

We investigate to the bugs that can only be revealed by \toolname under different scenarios. We especially pay attention to the bugs related to the requirement of valid inputs (\ie username and password in the Login scenario). We find that only several bugs in the Login scenario are due to this issue. Therefore, we believe that the advantages of \toolname over the baselines are flukily due to the integration of valid inputs, but due to the comprehensive integration of domain knowledge on app business logic from human testers.

In order to better understand the capability of \toolname, we have an investigation into the bug types that can be revealed only by Monkey/Stoat, by both Monkey/Stoat and \toolname, and only by \toolname. For the bugs revealed by Monkey/Stoat, we find that one typical bug is revealed in a rare requirement of account registration, which is not included in our EKG. The app requires the user to click on the user agreement and finish reading it. Monkey/Stoat occasionally clicks on the link and reveals such a bug. We go through all the 30 apps for the Register scenario, this requirement only appears in one app. One typical bug found only by both Monkey/Stoat and \toolname is caused by clicking on the login button with the null value of the email field. The app does not process such an exception and causes a mismatch of the primary key in the database. One typical bug only found by \toolname is that during the ticket booking process of FlightTicket scenario, the app crashes if the ticket holder name is left blank when the user finally pays. Monkey/Stoat fails because such a step is a follow-up operation of a complex operation sequence, and Monkey/Stoat cannot meet the prerequisite operation sequence to reach this step.

In summary, \toolname can effectively find distinct bugs that are not covered by the state-of-the-art automated testing tools for conducting testing with the integration of domain knowledge from human testers. \toolname can be viewed as a strong complement to traditional automated testing tools, especially in finding bugs in critical and common scenarios on different apps.

\subsection{Threats to Validity}

One main factor that may pose a threat is that our work focuses on crowdsourced test reports that are mainly written in Chinese. However, we eliminate such a threat by utilizing advanced NLP technologies, which can also process reports in English well. Besides, some keywords in English are included in the crowdsourced test reports we use, like ``Button'', ``TextField'', ``username'', \etc Therefore, we believe that language will not be an obstacle to our approach.

The apps and scenarios in our evaluation might be a threat. However, we have taken efforts to eliminate this. First, we refer to both the Google app store and a famous mobile app research benchmark, which can authentically reflect the popularity of commercial and open-source apps, respectively. We also follow specific criteria to ensure the scenarios are quite common. Second, we pay attention to the diversity of the scenarios. For Login and Register, these two scenarios are relatively short and simple in business logic, but the sub-scenarios are quite more diverse. The rest scenarios are more complex in business logic and always have a long event sequence, but not so diverse in sub-scenarios. Third, we want to highlight that the purpose of this paper is to solve the limitations of current approaches in \textit{common} scenarios, the business logic of which can be extracted from crowdsourced test reports.

Another threat may be that our experiment results are verified manually. In order to eliminate the negative effects, we invite experienced testing experts to conduct the result verification. We ask the experts to cross-validate the results independently, which further improves the reliability of the results. Therefore, we hold that all the experiment results are convincing.

Besides, the availability and quality of the test reports may affect the quality of the constructed EKGS. However, in our work, we do not depend on individual test reports to construct the EKGs. Instead, we use several crowdsourced test reports involving specific testing scenarios. This practice can mitigate the potential threat caused by some test reports that lack some information or with low quality. Different test reports focusing on the same testing scenarios can corroborate each other to build more robust EKGs for the testing scenarios.

Further, the report source might be a threat. However, we have take some efforts to mitigate this threat. First, the crowdsourced testing platform we use to collect reports is widely used in mobile app testing, and the user quantity shows its popularity. Second, the crowdsourced testing mechanism adopted by MoocTest is almost identical with other large-scale commercial platforms, like Global App Testing, Baidu Crowdtesting, Testin, \etc Therefore, we believe that the reports from MoocTest can show the generalizability of \toolname.

\section{Discussion}
\label{sec:dis}

As far as we can know, our approach is the first approach that combines the domain knowledge of human testers with scenario-based mobile app testing. Compared with traditional automated testing approaches, which explore the AUTs following specific strategies, we conclude three aspects of the characteristics of our approach. First, \textbf{the testing goal is different.} Current approaches mainly focus on improving the coverage of activities and widgets by traversing all possible activity transitions with widget operations. \toolname is the first approach to focus on ``testing scenarios'', which refers to one specific functionality with complete and self-consistent business logic. Second, \textbf{the testing granularity is different.} Traditional approaches conduct testing based on the whole app, and the exploration strategies take AUTs as a whole to cover as many activities and widgets as possible. However, \toolname conducts testing at testing scenario granularity, corresponding to testing cognition of human testers. Third, \textbf{the evaluation metrics are different.} Due to the above differences, \toolname cannot be evaluated with traditional evaluation metrics, like code coverage, widget coverage, or activity coverage. Consequently, we propose suitable evaluation metrics to evaluate \toolname from five aspects, correctness, reliability, accuracy, adequacy, and usefulness.

\section{Related Work}
\label{sec:rw}

\subsection{KG Application in Software Engineering}

Knowledge graph, as one of the most effective ways to organize knowledgeable information \cite{guan2022event}, has been applied to many topics in software engineering research. 
Nayak \etal \cite{nayak2020knowledge} propose a method to generate the knowledge graph from software documents that are primarily unstructured and sparse. 
Zhao \etal \cite{zhao2021brain} present a brain-inspired search engine assistant, named DeveloperBot, based on a knowledge graph, which aligns with the cognitive process of humans and has the capacity to answer complex queries with explainability. 
Zhang \etal \cite{zhang2020scheme} introduce a software security requirement acquisition scheme based on the knowledge graph. 
Jiang \etal \cite{jiang2021research} establish a high-quality medical Q\&A knowledge graph based on relevant professional knowledge in medical Q\&A research. 
Huang \etal \cite{huang2016enhancing} introduce a method to abstract class diagrams and to use the knowledge graph as an intermediate layer to analyze and deal with class diagrams. 
Zhang \etal \cite{zhang2021non} propose a function-oriented approach for eliciting non-functional requirements based on a domain knowledge graph. 

More specifically, the knowledge graph has a wide range of applications in the software testing domain. 
Cheng \etal \cite{cheng2022conflict} propose an approach to automatically infer Python compatible runtime environments with the domain knowledge graph.
Wang \etal \cite{wang2017construct} introduce a general approach to utilize the bug knowledge graph for bug resolution. 
Zhou \etal \cite{zhou2018intelligent} employ the knowledge graph technology for intelligent bug fixing, \ie to study effective search and recommendation techniques based on the knowledge graph for effective bug understanding, bug location, and bug resolution. 
Yang \etal \cite{yang2021test} construct a software testing knowledge graph based on the historical data of testing tasks. The approach is used to recommend usable and valuable test cases that are more likely to find defects in software. 
Liu \etal \cite{liu2020generating} mention a knowledge graph-based approach, APIComp, for generating the API comparison results. 
Yin \etal \cite{yin2021api} propose an API learning service that targets at problems faced by inexperienced developers and the service constructs an API knowledge graph to provide API-related learning resources. 
Guo \etal \cite{guo2020crowdsourced} develop the KARA approach to solve the problem of crowdsourced testing requirements generation, which uses a knowledge graph to enrich the description of the steps. 
Ke \etal \cite{ke2020interpretable} develop a test case recommendation method that can quickly find suitable test cases from a large number of test cases. 
Kwapong \etal \cite{kwapong2019knowledge} present a KG-based framework for web API recommendation.

The above studies all show the values of the KG when it is applied in software engineering tasks, and it can effectively store heterogeneous data and can provide a better reference for the software engineering tasks \cite{yang2021test}. These studies inspire us to utilize KG to organize the domain knowledge of human testers. This paper is the first work that applies EKG in scenario-based mobile app testing.

\subsection{Bug Report to Tests \& Test Transferring}

Test reproduction from bug reports is another important research direction in mobile app testing, which can help app developers better confirm the bug root cause by replaying the bug triggering operation sequence \cite{li2023towards}. ReCDroid \cite{zhao2019recdroid} \cite{zhao2022recdroid+} analyzes bug reports to reproduce bugs by using NLP technology to extract key steps from reports. It then compares the target widgets with the widgets extracted from the app source file to confirm action targets. Yakusu \cite{fazzini2018automatically} transforms user-reported bugs into test cases by extracting and identifying widgets from mobile app source code. It processes bug reproduction steps using NLP technology and matches widgets with GUI widgets through deep learning techniques. GIFDroid \cite{feng2022gifdroid} utilizes video information to assist bug reproduction. It identifies keyframes in videos that reproduce bugs and compares them with the actual apps in order to generate the reproduction steps. Such approaches target at reproducing known bugs to help confirm the root cause. They directly use single test reports to generate test cases that only fit the original app. However, different with these approaches, \toolname targets at utilizing multiple test reports to abstract the testing scenarios to help test generation on completely different app. The generalizability of \toolname is not the focus of the report-based test reproduction approaches.

Test migration among different apps (\ie test generation from tests of other apps) is an emerging topic in mobile app testing. Some studies \cite{el2010systematic} \cite{gomez2013reran} \cite{lam2017record} \cite{rau2018poster} \cite{liang2023rida} have shown that similarities between different apps can help reuse test cases, significantly reducing mobile app testing overhead. Behrang \etal \cite{behrang2018automated} propose an approach, AppTestMigrator, leveraging similarities between the app GUIs of related apps to adapt and reuse GUI test cases, thereby reducing app test case generation overhead. However, such studies are still trying to migrate existing test operation sequences to new apps, and the replay of test operation sequences should follow the recorded operation sequences. Instead, \toolname extracts knowledge from crowdsourced test reports, and integrate them into the EKGs, which are then used to guide the exploration of testing scenarios of new apps. Therefore, \toolname is a different approach with existing test transferring approaches.

\subsection{Image-Aided Software Testing}

Images contain rich information, and the rise of a large number of CV technologies has promoted the use of image information by researchers in software testing. It is important to recognize that a GUI image is different from a normal image. Instead of a traditional understanding of pixels, \cite{yu2021prioritize} argues that app screenshots should be viewed as a meaningful collection of GUI widgets, and DeepPrior is proposed to prioritize crowdsourced test reports through a deep understanding of images. After empirical evaluation of existing methods, Chen \etal \cite{chen2020object} propose a new method combining traditional CV technology and a DL model, improving the performance of GUI widget detection.

Several studies have been focused on the problem of locating and identifying and analyzing GUI widgets based on image understanding. Xiao \etal \cite{xiao2019iconintent} propose IconIntent, which utilizes CV technology to identify sensitive UI widgets. Yu \etal \cite{yu2021layout} utilize image understanding technologies to locate widgets during the test script record and replay process. Liu \etal \cite{liu2020owl} introduce OwlEye to detect screenshots and localize the buggy region in the UI. REMAUI \cite{nguyen2015reverse} leverages CV and OCR technologies to automatically infer mobile application UI elements from images. To help visually impaired users, Chen \etal \cite{chen2020unblind} develop a deep learning-based model to automatically predict image-based button labels.

GUI-based image understanding plays an important role in the field of mobile app testing \cite{chang2010gui}. Chen \etal \cite{chen2019automated} propose an automated cross-platform GUI code generation framework. \cite{beltramelli2018pix2code} presents pix2code, leveraging deep learning to generate code from a single input image. Moran \etal \cite{moran2018machine} introduce ReDraw to generate GUI code via the machine learning method. 
Chang \etal \cite{chang2010gui} present a novel approach to GUI testing which leverages CV algorithms to help GUI testers automate their tasks. 
Pan \etal \cite{pan2020gui} propose Meter, which uses CV technology to automatically repair GUI test scripts. Li \etal \cite{li2019humanoid} introduce Humanoid, using a deep neural network model to predict how humans choose actions to guide test input generation.
Sometimes, even in GUI design, these techniques play an important role. Chen \etal \cite{chen2018ui} translate UI design images into GUI skeleton by combining advances in computer vision and machine translation. Chen \etal \cite{chen2020wireframe} propose a DL-based UI design search engine, describing information in images that words cannot encapsulate. It is clear that researchers in the field of mobile apps realize this and explore the implementation of GUI-based image understanding in all directions.
Such studies indicate that automated approaches can imitate human perspectives to understand mobile apps. For \toolname, it is important to obtain the domain knowledge of human testers from app GUI analysis that can be organized and utilized to guide the scenario-based testing.

\section{Conclusion}
\label{sec:con}

Faced with the limitations of current automated testing approaches that the exploration ignores the business logic due to the lack of domain knowledge, we propose the novel approach \toolname to integrate the domain knowledge from human testers into the automated testing process with the assistance of app GUI understanding and EKG. \toolname realizes the information gathering and utilization for automated scenario-based mobile app testing via EKG construction from the understanding of test reports of human testers, including app GUI screenshots indicating the app behaviors and the textual descriptions describing the test events for specific test scenarios. The main information for EKG construction includes texts, widgets, GUI structures, \etc During the automated testing, EKGs can provide feedback on which widgets to be operated to complete the scenarios. \toolname is capable of testing different sub-scenarios by exploring different branches. Briefly speaking, \toolname pushes forward the research of scenario-aware automated testing by imitating human practices and can realize completely automated testing on target testing scenarios for the first time.

\section*{Acknowledgments}

The authors would like to thank the editors and anonymous reviewers for their time and comments.
This work is supported partially by the National Natural Science Foundation of China (62141215, 62272220, 62372228).

\bibliographystyle{IEEEtran}
\bibliography{main}

\begin{IEEEbiography}[{\includegraphics[width=1in,height=1.25in,clip,keepaspectratio]{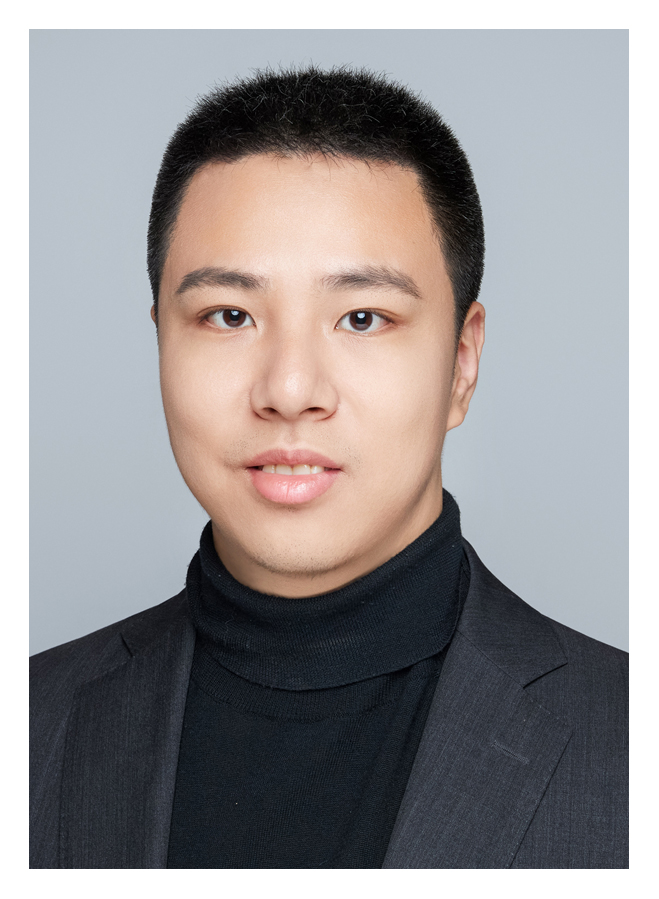}}]{Shengcheng Yu} (Student Member, IEEE) received the B.E. degree in software engineering from Software Institute, Nanjing University, Jiangsu, China. He is currently a Ph.D. candidate in software engineering at the Software Institute of Nanjing University. His research interests lie in software testing, GUI testing, crowdsourced testing, GUI understanding. He has published several research papers on venues including TSE, TOSEM, ICSE, FSE, \etc 
\end{IEEEbiography}

\begin{IEEEbiography}[{\includegraphics[width=1in,height=1.25in,clip,keepaspectratio]{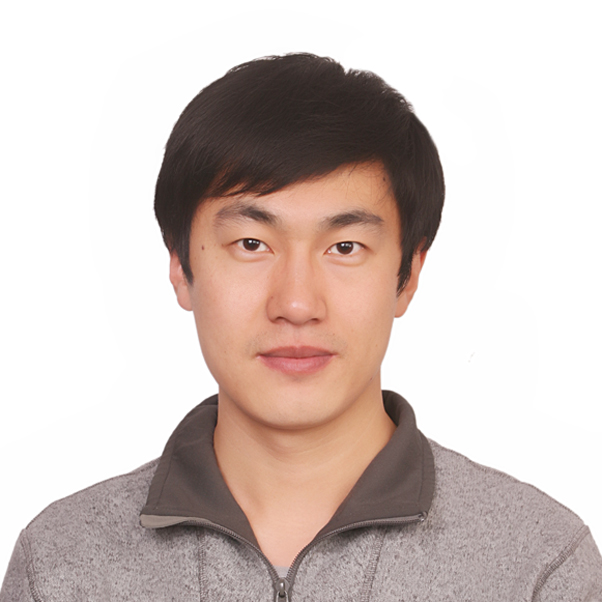}}]{Chunrong Fang} (Member, IEEE) received the B.E. and Ph.D. degrees in software engineering from Software Institute, Nanjing University, Jiangsu, China. He is currently an associate professor with the Software Institute of Nanjing University. His research interests lie in intelligent software engineering, e.g. BigCode and AITesting.
\end{IEEEbiography}

\begin{IEEEbiography}[{\includegraphics[width=1in,height=1.25in,clip,keepaspectratio]{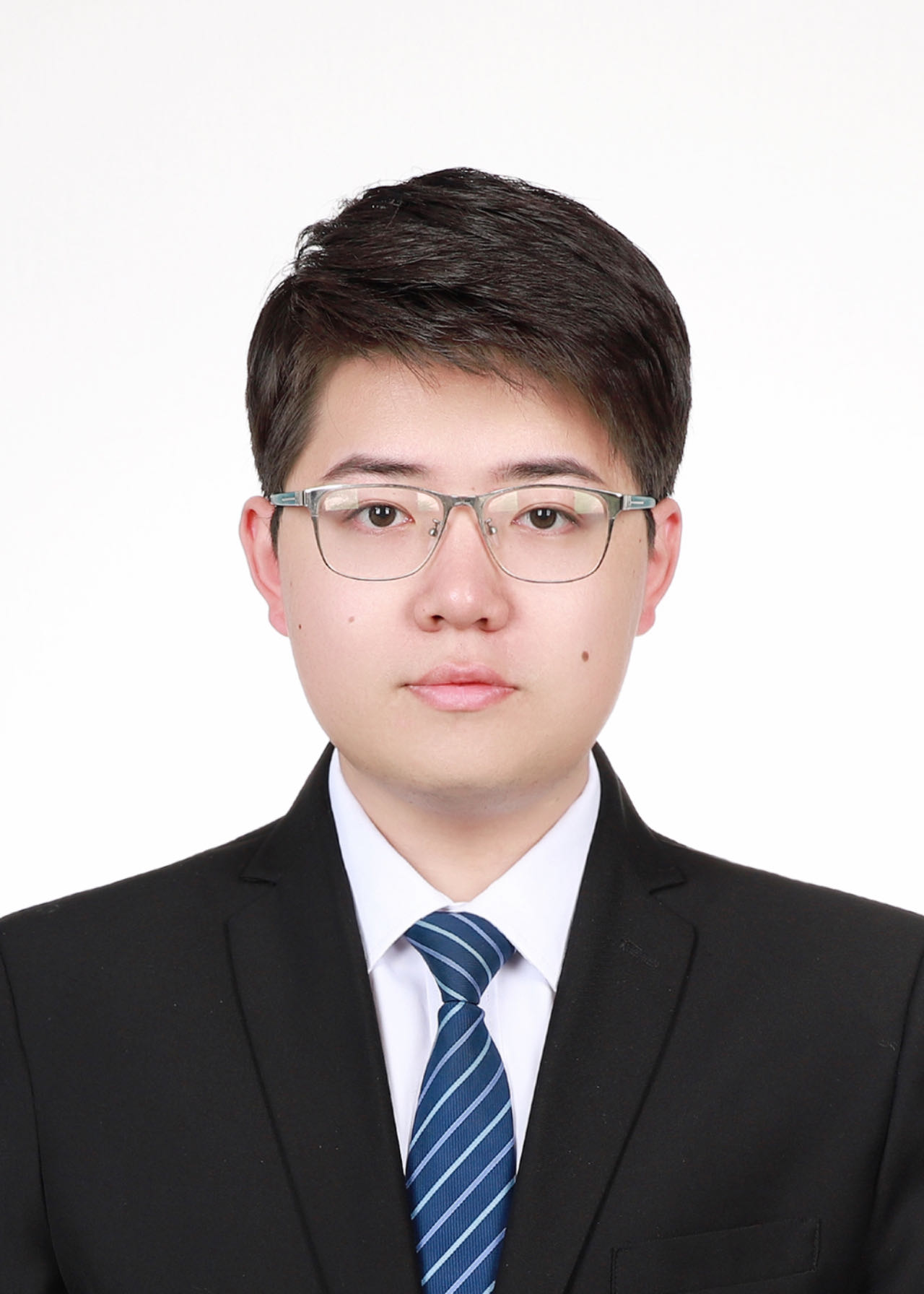}}]{Mingzhe Du} will be graduating from the Software Institute of Nanjing University, with a focus on mobile application testing during his master’s studies. He will join Ant Group as a software testing engineer.
\end{IEEEbiography}

\begin{IEEEbiography}[{\includegraphics[width=1in,height=1.25in,clip,keepaspectratio]{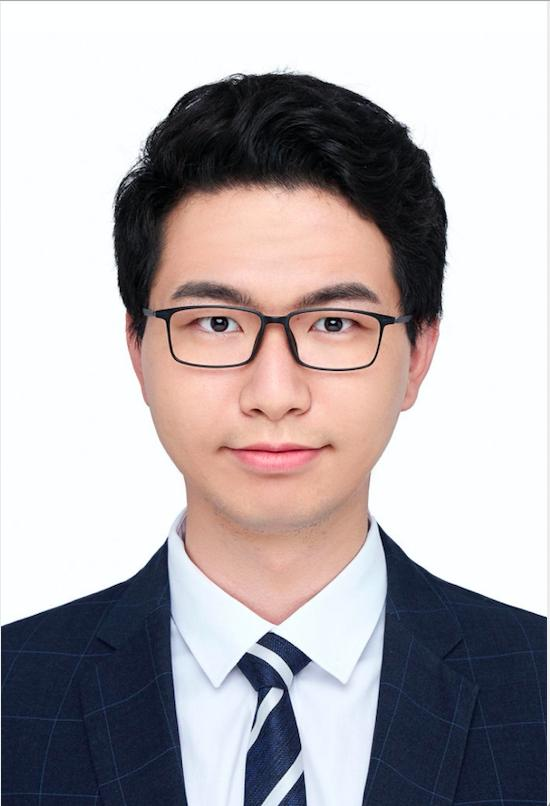}}]{Zimin Ding} is a postgraduate student at School of Software, Nanjing University, Nanjing, China. He received the B.E. degree in Computer Science and Technology from Southeast University, Nanjing, China. His current research interest is software testing.
\end{IEEEbiography}

\begin{IEEEbiography}[{\includegraphics[width=1in,height=1.25in,clip,keepaspectratio]{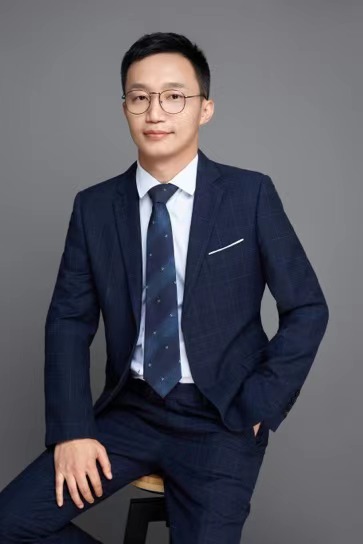}}]{Chenyu Chen} (Member, IEEE) is a full professor of Software Institute, Nanjing University, China. He is the main teacher of Statistical Methods and Data Analytics, and Software Testing: Methods and Techniques at Nanjing University, China. He has published a total of 86 papers as the first author or co-author. He is the associate Editor of IEEE Transactions on Reliability. He is also the Contest Co-Chair in China at QRS 2018, ICST 2019, and ISSTA 2019. Besides, he is the Industrial Track CoChair of SANER 2019, PC member of ISSRE 2018. His research interests include collective intelligence, deep learning testing and optimization, big data quality, and mobile application testing. 
\end{IEEEbiography}

\begin{IEEEbiography}[{\includegraphics[width=1in,height=1.25in,clip,keepaspectratio]{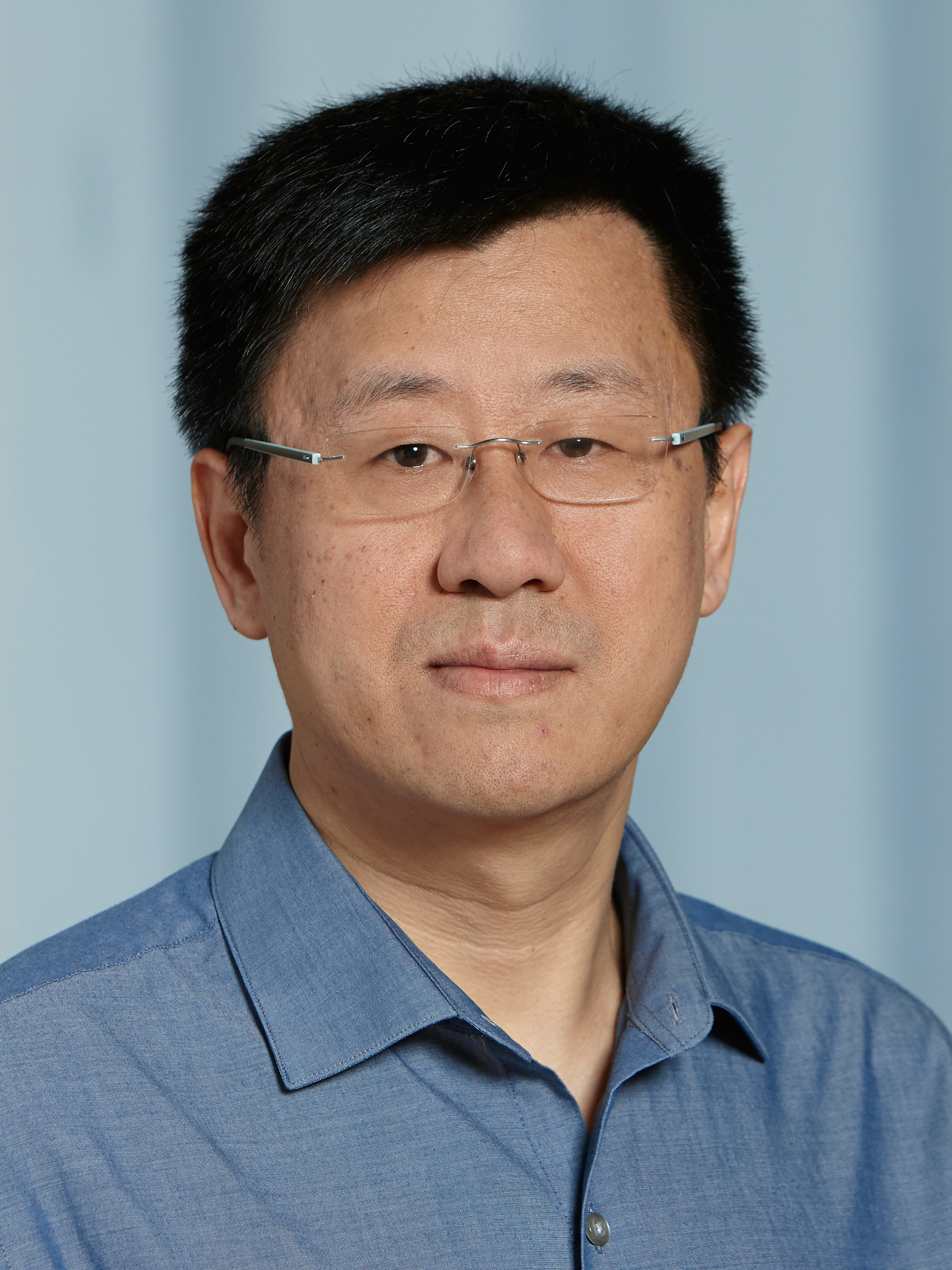}}]{Zhendong Su} (Fellow, IEEE) received a BS degree in computer science and a BA degree in mathematics from the University of Texas at Austin, Austin, Texas, and MS and PhD degrees in computer science from the University of California at Berkeley, Berkeley, California. He is a professor in computer science at ETH Zurich, Switzerland, where he specializes in programming languages and compilers, software engineering, computer security, deep learning, and education technologies. He is a member of the Academia Europaea and a fellow of the ACM. For more information, please visit https://people.inf.ethz.ch/suz/. 
\end{IEEEbiography}

\end{document}